\documentclass[10pt]{amsart}
\usepackage{amsfonts,amssymb}
\usepackage{amsmath}
\usepackage{color}
\usepackage{graphics}
\usepackage{epic}
\usepackage{curves}

\makeatletter
\@addtoreset{equation}{section}
\renewcommand{\theequation}{\thesection.\arabic{equation}}
\makeatother

\def\appendix#1{
\addtocounter{section}{1} \setcounter{equation}{0}
\renewcommand{\thesection}{\Alph{section}}
\section*{Appendix \thesection\protect\indent\quad
#1}
}

\catcode`\@=11
\def\marginnote#1{}

\newcount\hour
\newcount\minute
\newtoks\amorpm
\hour=\time\divide\hour by60 \minute=\time{\multiply\hour by60
\global\advance\minute by-\hour}
\edef\standardtime{{\ifnum\hour<12 \global\amorpm={am}%
        \else\global\amorpm={pm}\advance\hour by-12 \fi
        \ifnum\hour=0 \hour=12 \fi
        \number\hour:\ifnum\minute<10 0\fi\number\minute\the\amorpm}}
\edef\militarytime{\number\hour:\ifnum\minute<100\fi\number\minute}
\newcommand{\refpart}[1]{{\it (#1)}}

\def\draftlabel#1{{\@bsphack\if@filesw {\let\thepage\relax
      \xdef\@gtempa{\write\@auxout{\string
          \newlabel{#1}{{\@currentlabel}{\thepage}}}}}\@gtempa \if@nobreak
    \ifvmode\nobreak\fi\fi\fi\@esphack} \gdef\@eqnlabel{#1}}
    \def\@eqnlabel{}
\def\@vacuum{}
\def\draftmarginnote#1{\marginpar{\raggedright\scriptsize\tt#1}}

\def\draft{
  \oddsidemargin -.5truein
  \def\@oddfoot{\footnotesize \sl preliminary draft \hfil
    \rm\thepage\hfil\sl\today\quad\militarytime}
  \let\@evenfoot\@oddfoot \overfullrule 3pt
    \let\label=\draftlabel
    \let\marginnote=\draftmarginnote
  \def\@eqnnum{(\theequation)\rlap{\kern\marginparsep\tt\@eqnlabel}%
    \global\let\@eqnlabel\@vacuum}

  }

\newcommand{\CC}{\mathbb C}
\newcommand{\ZZ}{\mathbb Z}

\newcommand{\beq}{\begin{equation}}
\newcommand{\eeq}{\end{equation}}

\newcommand{\ID}{1\!\!1}

\def\be{\begin{equation}}
\def\ee{\end{equation}}
\def\bea{\begin{eqnarray}}
\def\eea{\end{eqnarray}}
\def\<{\langle}
\def\>{\rangle}

\def\nn{\nonumber}

\def\Tr{{\rm Tr}}
\def\one#1{#1^{\raise5pt\hbox{$\scriptstyle\!\!\!\!1$}}\,{}}
\def\two#1{#1^{\raise5pt\hbox{$\scriptstyle\!\!\!\!2$}}\,{}}
\def\onetwo#1{#1^{\raise5pt\hbox{$\scriptstyle\!\!\!\!\!{12}$}}\,{}}

\newcommand{\PP}{\mathbb P}
\newcommand{\thinf}{\theta_{\infty\!}}
\newcommand{\equal}{&\hspace{-7pt}=\hspace{-7pt}&}
\newcommand{\cequal}{&\hspace{-8pt}&}
\newcommand{\flarrow}[1]{\stackrel{\hspace{1pt}#1}{\xrightarrow{\hspace{21pt}}}}  

\newtheorem{theorem}{Theorem}[section]
\newtheorem{lm}[theorem]{Lemma}

\newtheorem{prop}[theorem]{Proposition}
\theoremstyle{definition}

\newtheorem{remark}[theorem]{Remark}

\theoremstyle{remark}

\allowdisplaybreaks

\begin{document}
\title[Cubic and quartic transformations of PVI ]
{Cubic and quartic transformations of  the sixth Painlev\'e  equation
in terms of Riemann--Hilbert correspondence}
\author{Marta Mazzocco}
\noindent\address{M.~Mazzocco, Loughborough University, Leicestershire LE11 3TU, UK.}
\noindent\email{m.mazzocco@lboro.ac.uk}
\author{Raimundas Vidunas}
\noindent\address{Raimundas Vidunas, Kobe University, Japan}

\maketitle

\begin{abstract}
A starting point of this paper is a classification of 
quadratic polynomial transformations of the monodromy manifold
for the $2\times2$ isomonodromic Fuchsian systems associated to the Painlev\'e VI equation. 
Up to birational automorphisms of the monodromy manifold, we find three transformations.
Two of them are identified as the action of known quadratic or quartic transformations 
 of the Painlev\'e VI equation. The third transformation of the monodromy manifold
 gives a new transformation of degree 3 of Picard's solutions of Painlev\'e VI.
\end{abstract}

\tableofcontents

\section{Introduction}

In this paper we consider the sixth Painlev\'e equation PVI  \cite{fuchs, Sch, Gar1}
\begin{eqnarray} \label{eq:P6}
\ddot q&=&{1\over2}\left({1\over q}+{1\over q-1}+{1\over q-t}\right) \dot q^2 -
\left({1\over t}+{1\over t-1}+{1\over q-t}\right)\dot q+\nn\\
&+&{q(q-1)(q-t)\over t^2(t-1)^2}\left[\alpha+\beta {t\over q^2}+
\gamma{t-1\over(q-1)^2}+\delta {t(t-1)\over(q-t)^2}\right],
\end{eqnarray}
as the isomonodromic deformation equation of a general linear $2\times2$ Fuchsian system
with 4 singularities $0,t,1,\infty$ (see Appendix \ref{se:iso}). Locally, solutions of the sixth Painlev\'e equation (up to Okamoto birational transformations) are in one-to-one correspondence, the so called {\it Riemann--Hilbert correspondence,}\/  with points on the monodromy manifold associated to such Fuchsian system. 

\begin{figure}
\[ 
\begin{picture}(270,180)
\put(0,0){\resizebox{270pt}{!}{\includegraphics{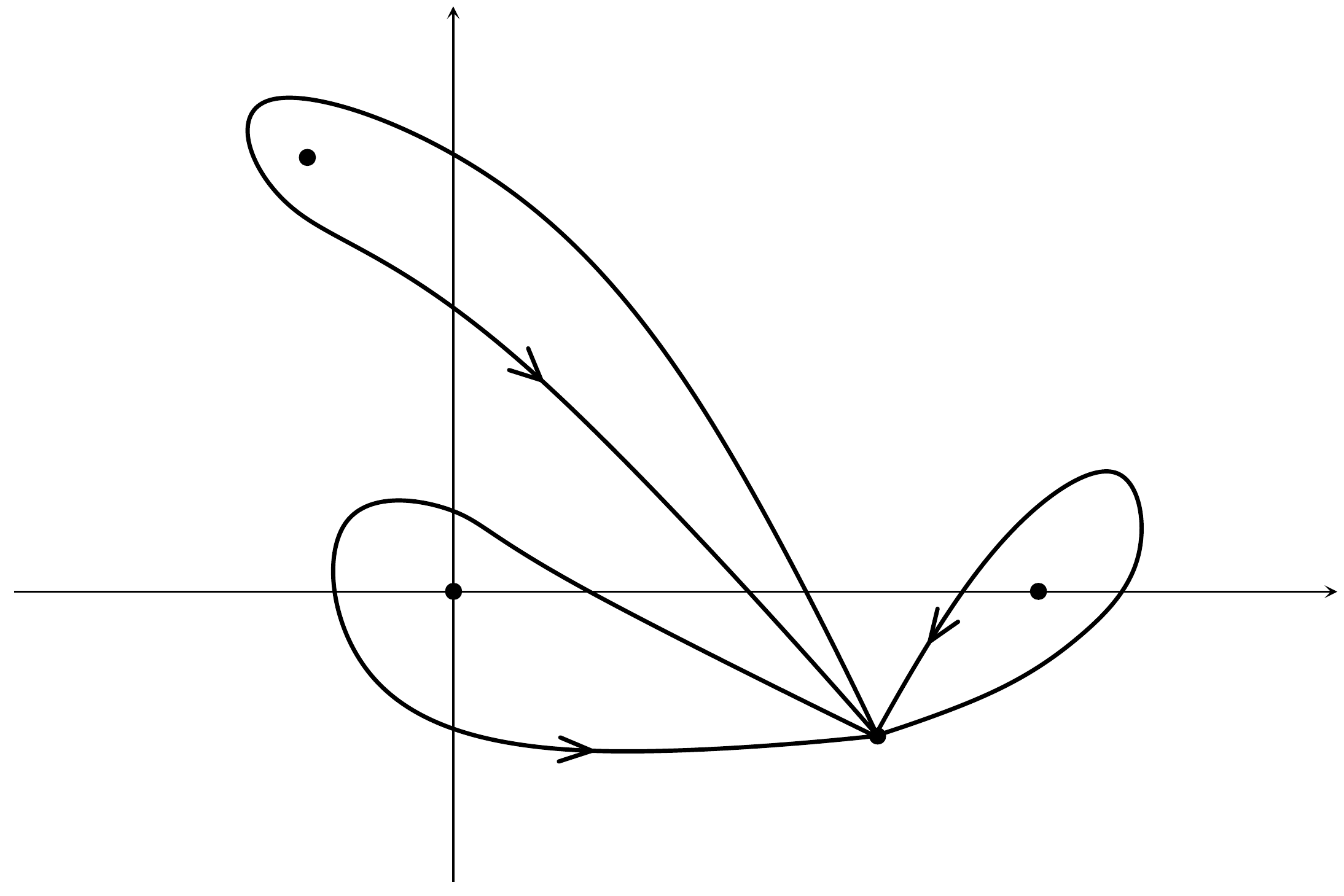}}}
\put(94,51){0} \put(206,51){1} \put(66,145){$t$}
\put(63,38){$\Gamma_0$}  \put(228,87){$\Gamma_1$}  
\put(113,138){$\Gamma_t$} \put(210,152){$\PP^1_{\nu}$}
\end{picture}
\] 
\caption{The basis of loops in $\mathbb P_\lambda$.}
\label{triangles237}
\end{figure}
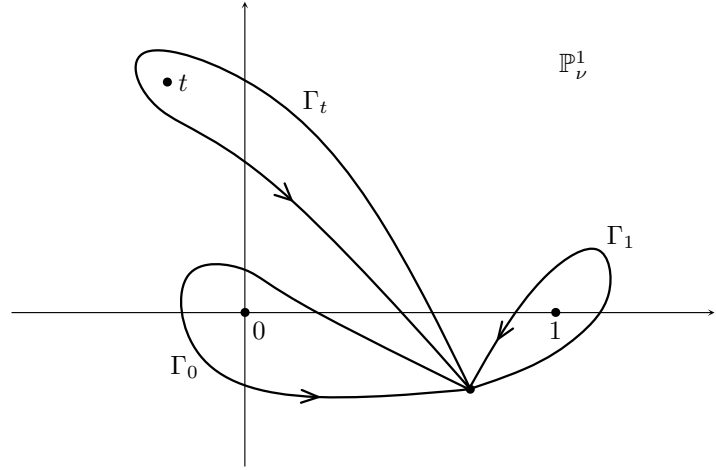

Let us briefly recall this setting. Denote by  $M_0,M_t,M_1,M_\infty\in SL(2,\mathbb C)$  the monodromy matrices  around the singularities $0,t,1,\infty$ with respect to the basis of loops depicted in Figure \ref{triangles237}.
Their traces are determined by the parameters
 $(\theta_0,\theta_t,\theta_1,\theta_\infty)$ such that
 \begin{equation}\label{eq:param}
\alpha=\frac{(\theta_\infty-1)^2}{2},\quad\beta=-\frac{\theta_0^2}{2},\quad \gamma=\frac{\theta_1^2}{2},
\quad \delta=\frac{1-\theta_t^2}{2}.
\end{equation}
Let ${\bf a}$ denote the vector $(a_0,a_t,a_1,a_\infty,a_{0t},a_{01},a_{t1})$ with
\begin{align} \label{eq:ai}
&a_i:= \Tr(M_i)= 2\cos(\pi \theta_i),\quad\mbox{for}\quad i\in\{0,t,1,\infty\}, \\
\label{eq:aij}
&a_{0t}= \Tr\left(M_0M_t\right),\quad a_{01}= \Tr\left(M_0M_1\right),
\quad a_{t1}= \Tr\left(M_tM_1\right).
\end{align}
As described in Appendix \ref{sec:rh},
the monodromy manifold  \cite{jimbo}
is represented by the following affine cubic surface in $\mathbb C^3$:
\be\label{eq:friecke}
Q:=\mathbb C[a_{0t},a_{01},a_{t1}]\slash\langle
a_{0t}^2+a_{01}^2+a_{t1}^2+a_{0t}a_{01}a_{t1}-\omega_{0t}a_{0t}-\omega_{01}a_{01}-\omega_{t1}a_{t1}+\omega_\infty=0\rangle.
\ee
Here $\omega_{0t},\omega_{01},\omega_{t1}$ and $\omega_{\infty}$ are given by
\begin{align} \label{eq:oij}
&\omega_{ij}:=a_i a_j + a_k a_\infty, \quad k\neq i, j,\quad \mbox{and} \quad i,j,k\in\{0,1,t\}, \\
&\omega_\infty = a_0^2+a_t^2+a_1^2+a_\infty^2+a_0 a_t a_1 a_\infty-4. \nn
\end{align}
The following birational automorphisms of the monodromy manifold are known:
\begin{itemize}
\item permutations of the coordinates $a_{0t},a_{01},a_{t1}$,
and the same permutation of the coefficients $\omega_{0t},\omega_{01},\omega_{t1}$;
\item changes of two signs, say 
$(\omega_{0t},\omega_{01},\omega_{t1})\mapsto(-\omega_{0t},-\omega_{01},\omega_{t1})$ 
and\\ $(a_{0t},a_{01},a_{t1})\mapsto(-a_{0t},-a_{01},a_{t1})$;
\item action of the braid group with these generators:
$$\beta_1(
a_{0t},a_{01},a_{t1}) = (
a_{0t},a_{t1},\omega_{01}-a_{01}-a_{0t}a_{t1}),$$
$$\beta_2(
a_{0t},a_{01},a_{t1})=(
a_{01},\omega_{0t}-a_{0t}-a_{01}a_{t1},a_{t1}).
$$
\end{itemize}
The first two correspond to Okamoto transformations, as recalled in the Appendix \ref{se:bir}, while the
braid group action describes analytic continuation of PVI solutions around the critical points \cite{DM,iwa}.

There are also other transformations acting on the solutions of the sixth Painle\'e equation:  quadratic and quartic transformations of the Painlev\'e VI equation are known \cite{kitaev,man,RGT,OST},
though the corresponding action on the monodromy manifold has not been presented
yet in the literature. This paper shows that these actions are given by quadratic polynomial transformations on the coordinates $a_{0t},a_{01},a_{t1}$ of the monodromy manifold.

More generally, we classify all quadratic polynomial transformations 
of the cubic surface (\ref{eq:friecke}) and identify the corresponding transformations of the sixth Painlev\'e equation. As a result, we find a new cubic transformation of the sixth Painlev\'e equation. Our classification result is summarised in the following theorem:

\begin{theorem}\label{th:class}
Up to the birational automorphisms of the monodromy manifold  (\ref{eq:friecke}),
the only transformations  of the form
$$
\tilde a_{0t}= X_1(a_{0t},a_{01},a_{t1}),
\qquad
\tilde a_{01}= X_2(a_{0t},a_{01},a_{t1}),
\qquad
\tilde a_{t1}= X_3(a_{0t},a_{01},a_{t1}),
$$
where $X_1,X_2,X_3$ are polynomials of degree 2 in $a_{0t},a_{01},a_{t1}$, which transform the cubic surface (\ref{eq:friecke}) with given parameters $\omega_{0t},\omega_{01},\omega_{t1}$ to a cubic surface of the same form with parameters  $\tilde\omega_{0t},\tilde\omega_{01},\tilde\omega_{t1}$
belong to the following list:
\begin{itemize}
\item Transformation mapping the cubic surface (\ref{eq:friecke}) with parameters $\omega_{0t}=\omega_{01}=0$ to the cubic surface (\ref{eq:friecke})  with parameters $\tilde\omega_{0t}=2\omega_{t1}$,
$\tilde\omega_{01}=\omega_\infty+4$, $\tilde\omega_{t1}=2\omega_{t1}$,
$\tilde\omega_\infty=\omega_{t1}^2+2\omega_\infty+4$:
$$
(a_{0t},a_{01},a_{t1})\mapsto(\omega_{t1}-a_{0t}a_{01}-a_{t1},2-a_{01}^2,a_{t1})
$$
\item Transformation mapping the cubic surface (\ref{eq:friecke}) with parameters $\omega_{0t}=\omega_{01}=\omega_{1t}=0$ to the  cubic surface (\ref{eq:friecke})  with parameters $\tilde\omega_{0t}=\tilde\omega_{01}=\tilde\omega_{t1}=2\omega_\infty+8$, 
$\tilde\omega_\infty=\omega_\infty^2+12 \omega_\infty+8$:
$$
(a_{0t},a_{01},a_{t1})\mapsto(2-a_{0t}^2,2-a_{01}^2,2-a_{t1}^2).
$$
\item The transformation mapping the cubic surface with parameters $\omega_{0t}=\omega_{01}=0$, $\omega_\infty=-4$ to itself: 
$$
(a_{0t},a_{01},a_{t1})\mapsto(-a_{0t}-a_{01}a_{t1},-a_{01}-a_{0t}a_{t1},-a_{t1}-a_{01}a_{0t}).
$$
\end{itemize}
\end{theorem}
The proof of this theorem is based on the properties
of the Poisson brackets (\ref{eq:gen-poisson}) on the monodromy manifold,
aided by some geometric insights and use of computer algebra.

The next set of results of this paper concerns the interpretation of each element in the list of 
Theorem \ref{th:class} in terms of Painlev\'e six transformations. We show that the first item in the list corresponds to a quadratic transformation, the second one corresponds to a quartic transformation, while the last item is a new cubic transformation of the Picard case of the sixth Painlev\'e equation.

Let us explain these results in more detail. 

We start from the quadratic transformations. Recall that quadratic transformations apply to the sixth Painlev\'e equation with restricted parameters. Here is the list of quadratic transformations that appeared in the literature:
\begin{eqnarray}
\hbox{Kitaev \cite{kitaev}:  }&\qquad
\left(\frac{1}{2},\theta_t,\theta_1,\pm\frac{1}{2}\right)\to
\left(\theta_1,\theta_t,\theta_1,\theta_t\right), &\label{eq:kit1}\\
\hbox{Manin \cite{man}:  }&\qquad
\left(\theta_1,\theta_t,\theta_1,\theta_t+1\right)\to
\left(2\theta_1,0,0,2\theta_t+1\right), &\label{eq:man}\\
\hbox{R.G.T. \cite{RGT}: }&\qquad
\left(\theta_1,\theta_t,\theta_t,\theta_1+1\right)\to
\left(0,2\theta_1,2\theta_t,1\right), &\label{eq:RGT}\\
\hbox{T.O.S. \cite{OST}: }&\qquad
\left(\theta_1,\theta_t,\theta_t,\theta_1+1\right)\to
\left(0,2\theta_t,0,2\theta_1+1\right), &\label{eq:TOS}
\end{eqnarray}
where R.G.T. is the abbreviation 
of Ramani, Grammaticos and Tamizhami, 
while T.O.S. is the abbreviation of Tsuda, Okamoto and Sakai. 
These transformations are all related among each other by Okamoto's  birational transformations.
(The fact that the last three are related by Okamoto symmetries is very easy to prove, the equivalence between the first one and the last three is a little more tricky and was carried out explicitly in \cite{FG}). 
Here we recall the explicit formula of the simplest (in our view), which is the T.O.S..:
\be
\big(q(t),t\big) \mapsto \displaystyle \big(\tilde{q}(\tilde{t}),\tilde{t}\big),
\qquad
\tilde q= \frac{(q+\sqrt{t})^2}{4q\sqrt{t}},
\quad\tilde t= \frac{(\sqrt{t}+1)^2}{4\sqrt{t}}.
\end{equation}
Note that the variable $t$ changes under this transformation. We prove the following:

\begin{theorem}\label{th:kit}
The Tsuda-Okamoto-Sakai 
transformation (\ref{eq:TOS}) acts on the vector ${\bf a}$ as follows:
\begin{eqnarray*}\label{eq:quartic} 
&&{\bf a}=(a_0,a_t,a_t,-a_0,a_{0t},a_{01},a_{t1})\to\\
&&
\qquad\to\tilde{\bf a}=(2,a_t^2-2,2,2-a_0^2,a_{t1},2-a_{01}^2,
a_t^2-a_0^2-a_{0t}a_{01}-a_{t1}).\nn
\end{eqnarray*}
\end{theorem}

Let us now concentrate on the quartic transformations. All quartic transformations known so far are related by Okamoto's birational canonical transformations to the folding transformation $\psi_{VI}^{[4]}$ due to Tsuda, Okamoto and Sakai  \cite{OST}:
$$
\begin{array}{rccc}
\psi^{(4)}_{VI}:&
 PVI(\vartheta,\vartheta,\vartheta,\vartheta+1)&\to& PVI(0,0,0,4\vartheta+1)\\
 &(q,t)&\to &\left(\tilde q,\tilde t\right)\\
 \end{array}
 $$
where $\tilde t=t$ and
$$
\tilde q= \frac{(q^2-t)^2}{4q(q-1)(q-t)}.
$$
We prove the following:

\begin{theorem}\label{th:q4}
The folding transformation  $\psi_{VI}^{[4]}$  acts on ${\bf a}$ as follows:
\begin{equation}\label{eq:qm}
{\bf a}=(a_0,a_0,a_0,-a_0,a_{0t},a_{01},a_{t1})\to
\tilde{\bf a}=(2,2,2,4a_0^2-2-a_0^4,2-a_{0t}^2,2-a_{01}^2,2-a_{t1}^2).
\end{equation}
\end{theorem}

This theorem can be proved as a Corollary of Theorem \ref{th:kit} by choosing the parameters $\theta_t$ and $\theta_1$in (\ref{eq:TOS}) in such a way that we can apply two quadratic transformations - up to Okamoto's birational canonical transformations.
Here, we present a more forthright  proof following Kitaev's \cite{kitaev} approach
of constructing a direct RS-pullback transformation on the isomonodromic Fuchsian system 
and deducing the transformation on the monodromy matrices (see Section  \ref{se:q4}). 

The reason for publishing this proof is that the transformation $a_{ij}\to 2-a_{ij}^2$ holds true
in a much more general setting and was used in \cite{ChMa} to show that the algebra of geodesic--length--functions on a disk with $n$ orbifold points coincides with the Dubrovin--Ugaglia algebra \cite{uga} of the Stokes data appearing in Frobenius Manifold theory.
Another advantage of this proof is that it gives a direct RS-transformation for the quartic
Painlev\'e VI transformation, simpler than a composition of two Kitaev's RS-transformations.

Picard's  case of the sixth Painlev\'e equation is given by
$\theta_0=\theta_t=\theta_1=0$, $\theta_\infty=1$, or equivalently,
$\alpha=\beta=\gamma=0,\delta=\frac{1}{2}$ in (\ref{eq:P6}).
In this special case, two quadratic transformations can be composed in an alternative way
to produce a degree 4 transformation.
\begin{prop} \label{th:altq4}
If $q(t)$ is a solution of Picard's case of the sixth Painlev\'e equation,
then $\tilde q\left(\tilde t\right)$ with
$$ 
\tilde q= -\frac{\left(\sqrt[4\,]t+1\right)^4 \, q \left(q+\sqrt{t}\right)^2}{(q-1)\,(q-t)\,(q-\sqrt{t})^2},
\quad\tilde t= \frac{\left(\sqrt[4\,]t+1\right)^4}{\left(\sqrt[4\,]t-1\right)^4}
$$
is a solution of the same Painlev\'e equation as well.
\end{prop}

As is known, Hitchin's case $\theta_0=\theta_t=\theta_1=\theta_\infty=\frac12$
of the sixth Painlev\'e equation \cite{hit}
is an Okamoto transformation of Picard's case. But the corresponding quadratic
and quadric transformations are more complicated for Hitchin's case; 
see Proposition \ref{th:vkit} to get an impression.
 
The last item in the list of Theorem \ref{th:class} gives a new {\it cubic}\/  transformation 
of the same Picard case. 
\begin{theorem}\label{th:cubic}
Let us parameterise the independent variable $t\in\mathbb C$ in terms of a new independent variable $s\in\mathbb C$ by imposing
$$
t=\frac{s^3(s+2)}{2s+1}.
$$
then the transformation $(q,t)\mapsto(\tilde q,\tilde t)$
$$
\tilde t= \frac{s(s+2)^3}{(2s+1)^3},\qquad \tilde q= \frac{q(q+s(s+2))^2}{((2s+1)q+s^2)^2}
$$
preserves the Picard case $\alpha=\beta=\gamma=0,\delta=\frac{1}{2}$ of the sixth Painlev\'e  equation.
\end{theorem}

Note that the $t$-variable changes under this transformation,
similarly as in Kitaev's (etc) quadratic transformation. 

This theorem is proved  using the explicit form of Picard's
solutions of PVI in terms of the Weierstrass $\wp$-function (Section \ref{se:Pic}). 
In fact, we prove that the action of the cubic transformation on the PVI solution $q(t)$
coincides  with an isogeny  of degree $3$ of the underlying Legendre's elliptic curves:
$$
w^2=q(q-1)(q-t).
$$
More precisely, we recall that an elliptic curve is a smooth, projective algebraic curve of genus one, on which there is a specified point  $\mathcal{O}$ \cite{silverm}. Points on the elliptic curve
have a structure of an abelian group (isomorphic to the Jacobian variety of the genus 1 curve),
and $\mathcal{O}$ is assigned to be the neutral element of the group. An {\em isogeny} between elliptic curves is an morphism between the genus $1$ curves that identifies the identity elements,
and is therefore a group homomorphism. The cubic isogeny is a group homomorphism with the kernel isomorphic to $\ZZ/3\ZZ$. The transformation between $t$ and $\tilde t$ identifies
the generic family of Legendre elliptic curves connected by a cubic isogeny. 
Over $\CC$, the period lattices of the isogenous curves are sublattices of each other (of index 3)
up to homothety. The action on the PVI solution of the cubic transformation coincides with
the action of the cubic isogeny on the $q$-coordinate of the Legendre curve.

It is clear that isogenies of any degree will act in the similar way. For example the quadratic transformations applied to Picard's case
correspond to degree $2$ isogenies, and the quadric transformation 
$\psi_{VI}^{[4]}$  corresponds to multiplication by $2$ map on the elliptic curves. 
The quartic transformation of Proposition \ref{th:altq4} corresponds to degree 4 isogenies
that are not multiplication by $2$ maps.

More generally, an isogeny of degree $n$ of Legendre's elliptic curve will produce a polynomial transformation of the monodromy data $(a_{0t},a_{01},a_{t1})$ on the Markov cubic:
$$
\mathbb C[a_{0t},a_{01},a_{t1}]\slash\langle
a_{0t}^2+a_{01}^2+a_{t1}^2+a_{0t}a_{01}a_{t1}-4\rangle.
$$ 
We conjecture that, apart from higher order isogenies, our list in Theorem \ref{th:class} is complete, without restricting the order of the polynomials $X_1,X_2,X_3$. This conjecture is suggested by the fact that as we go down the list, the parameters for which the transformations are defined become more and more specialised: for the quadratic transformations on needs to fix two parameters, for the quartic one three, for the cubic all four parameters need to be fixed. 

This paper is organised as follows: in Section \ref{se:cl} we prove our classification theorem. In Section \ref{se:kit} we prove Theorem \ref{th:kit}. In Section \ref{se:q4} we deal with the quartic transformation.  In 
Section \ref{se:Pic}  we discuss the Picard case of PVI and prove Theorem \ref{th:cubic}.
In Appendix A we remind a few facts about the isomonodromic deformation problem associated to the sixth Painlev\'e equation and the monodromy manifold. In Appendix B we recall Okamoto's birational transformations and their action on the monodromy manifold.

\vskip 2mm \noindent{\bf Acknowledgements.} The authors are grateful to Alexander Kitaev for his helpful suggestions and remarks, and to Leonid Chekhov, Davide Guzzetti, and Masatoshi Noumi for useful conversations. This research  was
supported by the EPSRC ARF EP/D071895/1 and JSPS 20740075.

\vskip 2mm

\section{Classification of quadratic transformations on the monodromy manifold}\label{se:cl}

In this section we prove theorem \ref{th:class}. The proof relies heavily on some properties of the Poisson structure (\ref{eq:gen-poisson}) on the monodromy manifold, which we recall in the next subsection.

\subsection{Poisson structure on the monodromy manifold}\label{ss:aut}

Given any polynomial $C\in\mathbb C[a_{0t},a_{01},a_{t1}]$, the following formulae define a Poisson bracket on
$\mathbb C[a_{0t},a_{01},a_{t1}]$:
\be\label{eq:EG}
\{a_{0t},a_{01}\} =\frac{\partial{C}}{\partial a_{t1}},\qquad \{a_{01},a_{t1}\}=\frac{\partial{C}}{\partial a_{0t}},
\qquad \{a_{t1},a_{0t}\}=\frac{\partial{C}}{\partial a_{01}},
\ee
and ${C}$ itself is a central element for this bracket, so that the quotient space
$$
Q:= \mathbb C[a_{0t},a_{01},a_{t1}]\slash_{\langle{C}=0\rangle}
$$
inherits the Poisson algebra structure \cite{EG}. In the case of the cubic $Q$ defined by (\ref{eq:friecke}), the natural Poisson brackets in the Painlev\'e six monodromy manifold are given by 
\begin{eqnarray}
&&
\left\{a_{0t},a_{t1}\right\}= a_{0t} a_{t1} +2 a_{01} -  \omega_{01},\nn\\
&&
\left\{a_{t1},a_{01}\right\}= a_{t1} a_{01} +2 a_{0t} -  \omega_{0t},\label{eq:gen-poisson}\\
&&
\left\{a_{01},a_{0t}\right\}=  a_{0t} a_{01} +2 a_{t1} - \omega_{t1}.\nn
\end{eqnarray}
Therefore, any polynomial transformation (\ref{eq:friecke}) of the form
$$
\tilde a_{0t}= X_1(a_{0t},a_{01},a_{t1}),
\qquad
\tilde a_{01}= X_2(a_{0t},a_{01},a_{t1}),
\qquad
\tilde a_{t1}= X_3(a_{0t},a_{01},a_{t1}),
$$
which preserves the monodromy manifold must also preserve the Poisson structure  (\ref{eq:gen-poisson}) up to a constant factor. This fact plays a key role in the proof of Theorem  \ref{th:class}.

\subsection{Proof of Theorem \ref{th:class}}

First of all, since we wish to classify all transformations up to birational automorphisms of the cubic (\ref{eq:friecke}), to simplify notations we 
call the variables $a_{01},a_{0t},a_{1t}$ by $x_1,x_2,x_3$, and analogously  the constants $\omega_{01},\omega_{0t},\omega_{1t}$ are called  $u_0,u_1,u_2,u_3$ while the new $\tilde{\omega}_{01},\tilde{\omega}_{0t},\tilde{\omega}_{1t}$ are called  $v_1,v_2,v_3$. In these new notations, the monodromy manifold (\ref{eq:friecke}) is defined by the polynomial
\begin{equation} \label{eq:cubicC}
C=x_1^2+x_2^2+x_3^2+x_1x_2x_3-u_1x_1-u_2x_2-u_3x_3+u_0.
\end{equation}
We are looking for the quadratic transformations of the general form:
\begin{align}
\label{eq:xX}
X_1=&\eta_1x_1^2+\eta_2x_1x_2+\eta_3x_2^2+\eta_4x_1x_3+\eta_5x_2x_3+\eta_6x_3^2 \nn\\
&+\eta_7x_1+\eta_8x_2+\eta_9x_3+\eta_0,\nonumber \\
X_2=&\kappa_1x_1^2+\kappa_2x_1x_2+\kappa_3x_2^2+\kappa_4x_1x_3
+\kappa_5x_2x_3+\kappa_6x_3^2+\nn\\
&+\kappa_7x_1+\kappa_8x_2+\kappa_9x_3+\kappa_0,\\
X_3=&\xi_1x_1^2+\xi_2x_1x_2+\xi_3x_2^2+\xi_4x_1x_3+\xi_5x_2x_3+\xi_6x_3^2\nn\\
&+\xi_7x_1+\xi_8x_2+\xi_9x_3+\xi_0 \nonumber
\end{align}
where $\eta_0,\eta_1,\dots,\eta_9$,  $\kappa_0,\kappa_1,\dots,\kappa_9$ are constants, 
that transform the Poisson bracket (\ref{eq:gen-poisson})
\begin{align}
\{x_1,x_2\}&=x_1x_2+2x_3-u_3,\nonumber\\
\{x_2,x_3\}&=x_2x_3+2x_1-u_1,\\
\{x_3,x_1\}&=x_1x_3+2x_2-u_2\nonumber
\end{align}
on the cubic surface $C=0$, to the Poisson bracket:
\begin{align} \label{eq:xxbr1}
\{X_1,X_2\}&=K\,(X_1X_2+2X_3-v_3),\\  \label{eq:xxbr2} 
 \{X_2,X_3\}&=K\,(X_2X_3+2X_1-v_1),\\  \label{eq:xxbr3}
\{X_3,X_1\}&=K\,(X_1X_3+2X_2-v_2)
\end{align}
with $K\neq 0$. 

Let us introduce the following polynomials of degree $4$:
\bea\nn
&&
E_1:=\sum_{i,j=1}^{3}
\frac{\partial X_1}{\partial x_i}  \frac{\partial X_2}{\partial x_j}\{x_i,x_j\}-K\,(X_1X_2+2X_3-v_3),\nn\\
&&
E_2:=\sum_{i,j=1}^{3}
\frac{\partial X_2}{\partial x_i}  \frac{\partial X_3}{\partial x_j}\{x_i,x_j\}-K\,(X_2X_3+2X_1-v_1),\nn\\
&&
E_3:=\sum_{i,j=1}^{3}
\frac{\partial X_3}{\partial x_i}  \frac{\partial X_1}{\partial x_j}\{x_i,x_j\}-K\,(X_1X_3+2X_2-v_2),\nn
\eea
where $X_1,X_2,X_3$ are given by (\ref{eq:xX}). These three polynomials must be identically zero functions on the cubic (\ref{eq:cubicC}).

A polynomial in $\CC[x_1,x_2,x_3]$ is zero on the cubic  (\ref{eq:cubicC}) if and only if
it is a polynomial multiple of $C$. Alternatively, its normal form with respect to a Gr\"obner 
basis (consisting only of the polynomial $C$) must have all its coefficients equal to zero.
To recognise zero functions we thus must divide in $\CC[x_1,x_2,x_3]$ by $C$ with respect
to a term order. Division with respect to a total degree order (or most generally, 
with $xyz$ as the leading term) has the following geometric interpretation.
The cubic  (\ref{eq:cubicC}) intersects the infinity of $\PP^3$ at the three lines:
\begin{equation}
L_1: x_1=0,\qquad L_2:x_2=0,\qquad L_3:x_3=0.
\end{equation}
These lines intersect each other at the 3 infinite points:
\begin{equation}
P_1:x_2=x_3=0,\qquad P_2:x_1=x_3=0,\qquad P_3:x_1=x_2=0.
\end{equation}
The highest degree terms of a polynomial $f\in\CC[x_1,x_2,x_3]$ that are not divisible
by $xyz$ determine the restrictions of $f$ onto the lines $L_1,L_2,L_3$. Particularly,
the  terms with $x_1^d,x_2^d,x_3^d$ with $d=\deg f$ give the values of $f$
at $P_1,P_2,P_3$. 

Typically, we reduce polynomial functions $f$ on $C$ of degree 4.
The first steps of our division process are:
\begin{itemize}
\item Step P4: we require that the values of $f$ at the points $P_1,P_2,P_3$ to be zero;
\item Step L4: we require that  the restrictions of $f$ to the lines $L_0,L_1,L_2$ must zero functions; 
\item Step D4: we reduce the degree of $f$ by 1 with a single division
as all highest degree terms are divisible by $xyz$.
\end{itemize}
After performing these three steps,  the degree of $f$ is reduced to (at most) $3$. 
To reduce $f$ further to a quadratic or linear polynomial,
we repeat the same three steps as described,
but refer to them as P3, L3, D3, respectively, to stress that the highest degree is $3$.

We want to produce a classification up to the following symmetries:
\begin{enumerate}
\item[\refpart{$S_1$}] the cyclic permutations of $(x_1,x_2,x_3)$;
\item[\refpart{$S_2$}]  an odd permutation of $(x_1,x_2,x_3)$ combined 
with an odd permutation of $(X_1,X_2,X_3)$;
\item[\refpart{$S_3$}]  an odd permutation of $(x_1,x_2,x_3)$ or 
an odd permutation of $(X_1,X_2,X_3)$,  (in this case  the sign of $K$ changes); 
\item[\refpart{$S_4$}]  the sign change of two variables within $\{x_1,x_2,x_3\}$ or $\{X_1,X_2,X_3\}$.
\end{enumerate}
%
Here are the basic facts that help us to simplify our classification search.

\begin{lm}\label{lm:m0}
If two of the $X_j$'s are linear then $(X_1,X_2,X_3)$ is
either a permutation of $(x_1,x_2,x_3)$  or a braid group transformation.
\end{lm}

\proof
Let us assume that $X_1,X_2$ are linear. 
We can express $X_3$ using (\ref{eq:xxbr1}). Then
(\ref{eq:xxbr2})--(\ref{eq:xxbr3}) become
\begin{align} \label{eq:xxbr1a}
\{\{X_1,X_2\},X_2\}&=K^2\left(X_1X_2^2-4X_1-v_3X_2+2v_1\right),\\  \label{eq:xxbr2a}
\{X_1,\{X_1,X_2\}\}&=K^2\left(X_1^2X_2-v_3X_1-4X_2+2v_2\right).
\end{align}
Since the left-hand sides here
do not have the $x_1^3,x_2^3,x_3^3$ terms,  the coefficients
$\eta_7\kappa_7=\eta_8\kappa_8=\eta_9\kappa_9$ must be $0$. 
We assume that two linear coefficients of $X_1$ are zero, say $\eta_8=\eta_9=0$. 
Then $\eta_7\neq0$, $\kappa_7=0$, and (\ref{eq:xxbr2a}) has the terms
\[
(K^2-1)\eta_7^2x_1^2(\kappa_8x_2+\kappa_9x_3)+K^2\eta_7^2\kappa_0x_1^2+2K^2\eta_0x_1(\kappa_8x_2+\kappa_9x_3)+\ldots,
\]
and no $xyz$ term. Since we do not want both $\kappa_8,\kappa_9$ to be zero, 
we have $K=\pm 1$ and $\eta_0=\kappa_0=0$.  We adjust the (\ref{eq:xxbr1a}) with $\eta_7\kappa_8\kappa_9C$
to kill the $xyz$ term, and get the coefficient to $x^2$ equal to $(K^2+2)\eta_7\kappa_8\kappa_9$.
Hence $\kappa_8\kappa_9=0$, and $X_1,X_2$ are symmetric. We use this symmetry to assume $K=1$,
and then $\kappa_8=0$ leads to the displayed braid group transformation
with $(v_1,v_2,v_3)=(u_1,u_3,u_2)$, while $\kappa_9=0$ leads to the identity transformation.
\endproof

\begin{lm}\label{lm:m1}
Let us consider the matrix
\begin{equation}
M_1=\left( \begin{array}{ccc} \eta_1 & \eta_3 & \eta_6 \\
\kappa_1 & \kappa_3 & \kappa_6 \\ \xi_1 & \xi_3 & \xi_6 \end{array} \right)
\end{equation}
\begin{itemize}
\item In any column of the matrix $M_1$ there is at most one non-zero entry.
\item In any row of the matrix $M_1$ there is at most one non-zero entry.
\end{itemize}
\end{lm}

\proof
Step P4 gives the following  terms of degree $4$ in each variable $x_1,x_2,x_3$ of $E_1,E_2,E_3$:
\begin{align*}
E_1&=K\eta_1\kappa_1x_1^4+K\eta_3\kappa_3x_2^4+K\eta_6\kappa_6x_3^4+\ldots,\\
E_2&=K\kappa_1\xi_1x_1^4+K\kappa_3\xi_3x_2^4+K\kappa_6\xi_6x_3^4+\ldots,\\
E_3&=K\eta_1\xi_1x_1^4+K\eta_3\xi_3x_2^4+K\eta_6\xi_6x_3^4+\ldots.
\end{align*}
The displayed coefficients must be zero, and the first claim follows.

To show the last statement, let us assume by contradiction  that two entries in the first row
are non-zero, $\eta_1\neq0$, $\eta_3\neq0$. Therefore $\kappa_1=\kappa_3=\xi_1=\xi_3=0$ 
by the first statement. 
Here are some relevant terms for Step L4: 
\begin{align*}
E_1=&(K+2)\eta_3\kappa_2x_1x_2^3+(K+2)\eta_1\kappa_4x_1^3x_3+\big(K\eta_4\kappa_4+
(K+4)\eta_1\kappa_6\big)x_1^2x_3^2 
\\ &+(K-2)\eta_1\kappa_2x_1^3x_2+(p-2)\eta_3\kappa_5x_2^3x_3+\ldots,\\
E_3=&(K+2)\eta_1\xi_2x_1^3x_2+(K+2)\eta_3\xi_5x_2^3x_3+\big(K\eta_5\xi_5+(K+4)\eta_3\xi_6\big)x_2^2x_3^2 
\\&+(K-2)\eta_3\xi_2x_1x_2^3+(K-2)\eta_1\xi_4x_1^3x_3+\ldots.
\end{align*}
We still have the freedom of permuting $X_2,X_3$, so we assume $\Re(K)\ge0$. Then we
immediately have $\kappa_2=\kappa_4=\xi_2=\xi_5=0$, and then $\kappa_6=\xi_6=0$.
If $K\neq 2$, then $\kappa_5=\xi_4=0$ and $X_2,X_3$ are linear in $x_1,x_2,x_3$ thus leading to the contradiction that $\eta_1,\eta_3=0$.
Hence $K=2$. Then $E_2-\kappa_5\xi_4x_3C$ is of degree 3, and the coefficient to $x_3^3$ is $\kappa_5\xi_4$.
The variables $\kappa_5,\xi_4$ are still symmetric by \refpart{$S_2$}, so we assume $\kappa_5\neq0,\xi_4=0$.
Now $E_3$ is ready for Step P3: it has the terms
$2\kappa_5x_2x_3^2(\eta_5x_2+2\eta_6x_3)$, giving $\eta_5=\eta_6=0$.
We now perform Step 1 on $E_3$:
\[
E_3=2\eta_1\xi_7x_1^3+2\eta_3\xi_8x_2^3+4\eta_3\xi_9x_2^2x_3+\ldots.
\]
Hence $\xi_7=\xi_8=\xi_9=0$, and $X_3$ reduces to a trivial constant, which is not allowed.
\endproof

Up to the symmetries, we can assume that all non-zero entries of $M_1$ lie on the main diagonal.
Lemma \ref{lm:m1} then sets $\eta_3=\eta_6=\kappa_1=\kappa_6=\xi_1=\xi_3=0$,
and the polynomials $E_1,E_2,E_3$ are ready for Step L4:
\begin{align} \label{eq:eee}
\hspace{40pt} E_1=& (K+2)(\eta_1\kappa_4x_1^3x_3+\eta_5\kappa_3x_2^3x_3)+(K-2)(\eta_1\kappa_2x_1^3x_2+\eta_2\kappa_3x_1x_2^3) 
\nonumber \\&
+K(\eta_4\kappa_4x_1^2x_3^2+\eta_5\kappa_5x_2^2x_3^2)+\big(K\eta_2\kappa_2+(K-4)\eta_1\kappa_3\big)x_1^2x_2^2
+\ldots, \qquad\nonumber\\
E_2=& (K+2)(\kappa_3\xi_2x_1x_2^3+\kappa_4\xi_6x_1x_3^3)+(K-2)(\kappa_3\xi_5x_2^3x_3+\kappa_5\xi_6x_2x_3^3) \\&
+K(\kappa_2\xi_2x_1^2x_2^2+\kappa_4\xi_4x_1^2x_3^2)+\big(K\kappa_5\xi_5+(K-4)\kappa_3\xi_6\big)x_2^2x_3^2
+\ldots, \nonumber\\
E_3=& (K+2)(\eta_1\xi_2x_1^3x_2+\eta_5\xi_6x_2x_3^3)+(K-2)(\eta_1\xi_4x_1^3x_3+\eta_4\xi_6x_1x_3^3) 
\nonumber \\&
+K(\eta_2\xi_2x_1^2x_2^2+\eta_5\xi_5x_2^2x_3^2)+\big(K\eta_4\xi_4+(K-4)\eta_1\xi_6\big)x_1^2x_3^2
+\ldots. \nonumber
\end{align}
We split the proof into a few cases, and indicate a computational path to the results
of Theorem  \ref{th:class}.

\subsubsection{At least two entries of $M_1$ are non-zero.}

We assume here that $\eta_1\kappa_3\neq0$, but nothing immediately about $\xi_6$.
We distinguish the following cases:
\begin{itemize}
\item $K=-2 \implies$ from $E_1$ we have $\eta_2=\kappa_2=0$.\\
But then $(K-4)\eta_1\kappa_3=0$, a contradiction.
\item $K=2 \implies  \eta_5=\kappa_4=0$, $\eta_2\kappa_2=\eta_1\kappa_3\neq 0$ by $E_1$.
\item $K\neq\pm2\implies \eta_2=\eta_5=\kappa_2=\kappa_4=0$, 
$K=4$ by $E_1$, \\
and $\xi_2=\xi_4=\xi_5=0$, $\eta_4\xi_6=\kappa_5\xi_6=0$ by $E_2,E_3$.
\end{itemize}
In the second subcase, it is enough to work with $E_1$ to reach a contradiction.
Particularly, Steps D4 and P3 give 
$$
\eta_4\kappa_5=0, \quad 
2\eta_1(\kappa_5-\kappa_7)=\eta_4\kappa_2, \quad 
2\kappa_3(\eta_4-\eta_8)=\eta_2\kappa_5.
$$
The variables $\eta_4,\kappa_5$ are still symmetric by \refpart{S3}, 
so we assume $\kappa_5=0$. 
At Step P3 we consider 
$E_1-\eta_4(\kappa_2x_1+2\kappa_3x_2)C=-6\eta_4\kappa_3x_2x_3^2+\ldots$,
hence $0=\eta_4=\kappa_7=\eta_8$. 
Still at Step P3, we conclude $0=\kappa_8=\eta_7$, $\eta_9=\eta_2$, $\kappa_9=\kappa_2$.
But Step D3 gives $E_1+4\eta_1\kappa_3C=6\eta_1\kappa_3x_3^2+\ldots$, 
contradicting $\eta_1\kappa_3\neq 0$.

In the last subcase, Step D4 gives
\begin{align*}
E_2=&4\kappa_3\xi_8x_2^3+4\kappa_9\xi_6x_3^3+(5\kappa_5\xi_8+2\kappa_3\xi_9)x_2^2x_3\\
&+(2\kappa_8\xi_6+3\kappa_5\xi_9)x_2x_3^2+2\xi_6(3\kappa_7-2\kappa_5)x_1x_3^2+\ldots,\\
E_3=&4\eta_1\xi_7x_1^3+4\eta_9\xi_6x_3^3+(5\eta_4\xi_7+2\eta_1\xi_9)x_1^2x_3\\
&+(2\eta_7\xi_6+3\eta_4\xi_9)x_1x_3^2+2\xi_6(3\eta_8-2\eta_4)x_2x_3^2+\ldots.
\end{align*}
Step P3 gives $\xi_7=\xi_8=0$, and then Step L3 sets $\xi_9=0$.
We have to assume $\xi_6\neq 0$, as otherwise $X_3$ is a constant.
Therefore $\eta_4=\kappa_5=0$, and then $0=\kappa_9=\kappa_8=\kappa_7$
and $0=\eta_9=\eta_7=\eta_8$. There are no linear terms in the $X_j(x_1,x_2,x_3)$'s thus.
In Step D3 we compute the expressions $E_1+8\eta_1\kappa_3C$, $E_2+8\kappa_3\xi_6C$
$E_3+8\eta_1\xi_6C$. Their coefficients give the equations
\begin{align*}
u_1=u_2=u_3=0,\quad 
\eta_0=-2\eta_1,\quad \kappa_0=-2\kappa_3,\quad \xi_0=-2\xi_6,\\
\eta_1+\kappa_3\xi_6=0,\quad \kappa_3+\eta_1\xi_6=0,\quad \xi_6+\eta_1\kappa_3=0, 
\end{align*}
etc. This already implies that $\eta_1,\kappa_3,\xi_6\in\{1,-1\}$ and $\eta_1\kappa_3\xi_6=-1$.
Up to the symmetries, we have $\eta_1=\kappa_3=\xi_6=-1$, $\eta_0=\kappa_0=\xi_0=2$ and
eventually $v_1=v_2=v_3=2u_0+8$. We get the second transformation of Theorem \ref{th:class}.

\subsubsection{One non-zero entry of $M_1$}

Here we assume that $\eta_1\neq 0$ and all the other entries of $M_1$ equal to zero.

If $\kappa_2=\kappa_4=\xi_2=\xi_4=0$, then we assume $\kappa_5\neq0$ 
since we want $X_2$ or $X_3$ to have a quadratic term by Lemma \ref{lm:m0}. 
Then $\eta_5=\xi_5=0$, and we can assume \mbox{Re $K\ge0$}.
Step D4 reduces $E_1$ to
\begin{align*}
&-(p-3)\eta_2\kappa_5x_2^3-(p+3)\eta_4\kappa_5x_3^3+((p+1)\eta_4\kappa_9-(p+4)\eta_1\kappa_5)x_2x_3^2+\ldots.
\end{align*}
With the assumption Re $K\ge0$, we have $\eta_4=0$ and then $\eta_1\kappa_5=0$,
contradictorily.

Therefore we assume that at least one of the variables $\kappa_2, \kappa_4, \xi_2, \xi_4$
is non-zero. These variables are symmetric by \refpart{S3} and \refpart{S4}, 
so we assume $\kappa_2\neq 0$.
Then from (\ref{eq:eee}) we have $K=2$,  \mbox{$\kappa_4=\eta_2=\xi_2=0$}, $\eta_4\xi_4=0$,
and at most one of the variables $\eta_5,\kappa_5,\xi_5$  can be non-zero.
Step D4 reduces $E_1,E_2$ to 
\begin{align*}
&-5\eta_5\kappa_2x_2-5\eta_4\kappa_5x_3^3
+(2\eta_1\kappa_7-2\eta_1\kappa_5+\eta_4\kappa_2)x_1^3
+\kappa_2(\eta_7-\eta_5)x_1^2x_2+\ldots,\\
&-5\kappa_2\xi_4x_1^3+\kappa_2\xi_5x_2^3+\kappa_5\xi_4x_3^3
+3\kappa_2(\xi_7-\xi_5)x_1^2x_2+\kappa_2(\xi_8-\xi_4)x_1x_2^2+\ldots.
\end{align*}
Hence $\eta_5=\eta_7=0$, $\xi_4=\xi_5=\xi_7=\xi_8=0$. We still have cubic terms left
in 
\begin{align*}
E_1&-\big((2\eta_1\kappa_5+\eta_4\kappa_2)x_1+3\eta_4\kappa_5x_3\big)C=\\
&-5\eta_4\kappa_5x_3^3+\kappa_5(\eta_8-\eta_4)x_2^2x_3+3\eta_9\kappa_5x_2x_3^2
+(2\eta_1\kappa_5+3(\eta_8-\eta_4)\kappa_2)x_1x_2^2+\ldots
\end{align*}
and $E_2=\kappa_5\xi_9x_2x_3^2+\ldots$.
If $\kappa_5\neq0$, then $\eta_4=\eta_8=\eta_9=\xi_9=0$, yet the coefficient to $x_1x_2^2$
contradicts $\eta_1\kappa_5=0$. Therefore $\kappa_5=0$, $\eta_8=\eta_4$. 
Step D3 gives 
\begin{align*}
E_2-2\kappa_2\xi_9C=&4(\eta_1-\kappa_2\xi_9)x_1^2+2\xi_9(\kappa_9-\kappa_2)x_3^2+2\kappa_2\xi_0x_1x_2+\kappa_8\xi_9x_1x_3\\
&+(4\eta_4+3\kappa_7\xi_9)x_1x_3+\ldots.
\end{align*}
Hence $\xi_9\neq0$, $\eta_1=\kappa_2\xi_9$, $\kappa_9=\kappa_2$, $\xi_0=\kappa_8=0$. Then
\begin{align*}
E_3=\eta_4\xi_9x_1x_3^2+2\eta_9\xi_9x_3^2-4\kappa_2(\xi_9^2-1)x_1x_2+\ldots.
\end{align*}
This gives $\eta_4=\kappa_7=\eta_9=0$, $\xi_9=\pm 1$. We may assume $\xi_9=1$ by symmetry \refpart{S5}.
Then we have only linear terms left from $E_2,E_3$,
and conclude $u_1=u_2=0$,  $\kappa_0=-\kappa_2u_3$, $\eta_0=-2\kappa_2$.
After this $E_1=4(1-\kappa_2^2)x_3+4\kappa_2^2u_3-2v_3.$
We can take $\kappa_2=-1$ still by symmetry \refpart{S5}, and finalize 
$v_1=u_0+4$, $v_2=2u_3$, $v_3=2u_3$. We get the first 
transformation $(x_1,x_2,x_3)\mapsto (2-x_1^2,u_3-x_3-x_1x_2,x_3)$
of Theorem \ref{th:class}, from the cubic 
$x_1^2+x_2^2+x_3^2+x_1x_2x_3-u_3x_3+u_0=0$ to the cubic\\
$x_1^2+x_2^2+x_3^2+x_1x_2x_3-(u_0+4)x_1-2u_3x_2-2u_3x_3+u_3^2+2u_0+4=0$.


\subsubsection{$M_1$ is the zero matrix}

Only the quartic terms of $E_1,E_2,E_3$ with the factor $K$ in (\ref{eq:eee}) are non-zero.
We conclude the matrix
$$M_2=\left( \begin{array}{ccc} \eta_2 & \eta_4 & \eta_5 \\
\kappa_2 & \kappa_4 & \kappa_5 \\ \xi_2 & \xi_4 & \xi_5 \end{array} \right).$$
can have at most one non-zero entry in each column.
We want at least two rows of $M_2$ to be non-zero.
By symmetries, we assume that $\eta_5\neq 0$, $\kappa_4\neq 0$. 
Therefore $\eta_4=\kappa_5=\xi_4=\xi_5=0$.
Step D4 reduces E1 to
\begin{align*}
-(K+3)\eta_2\kappa_4x_1^3-(K+3)\eta_5\kappa_2x_2^3+(3-K)\eta_5\kappa_4x_3^3+\ldots.
\end{align*}
From here, $p=3$ and $\eta_2=\kappa_2=0$. For Step L3 we have:
\begin{align*}
E_1-2\eta_5\kappa_4x_3C &=2\eta_9\kappa_4x_1x_3^2+2\eta_5\kappa_9x_2x_3^2
+4\kappa_4(\eta_7-\eta_5)x_1^2x_3+4\eta_5(\kappa_8-\kappa_4)x_2^2x_3+\ldots,\\
E_2-2\kappa_4\xi_2x_1C &=2\kappa_7\xi_2x_1^2x_2+2\kappa_4\xi_7x_1^2x_3
+4\xi_2(\kappa_8-\kappa_4)x_1x_2^2+4\kappa_4(\xi_9-\xi_2)x_1x_3^2+\ldots,\\
E_3-2\eta_5\xi_2x_2C &=2\eta_8\xi_2x_1x_2^2+2\eta_5\xi_8x_2^2x_3
+4\xi_2(\eta_7-\eta_5)x_1^2x_2+4\eta_5(\xi_9-\xi_2)x_2x_3^2+\ldots.
\end{align*}
This gives $\eta_9=\kappa_9=\xi_7=\xi_8=0$, 
$\eta_7=\eta_5$, $\kappa_8=\kappa_4$, $\xi_9=\xi_2$. 
Step D3 reduces $E_1$ to the quadratic expression
\begin{align*}
-5\eta_8\kappa_4x_1^2-5\eta_5\kappa_7x_2^2+(\eta_5\kappa_4u_3-\eta_8\kappa_4-\eta_5\kappa_7)x_3^2+2(3\xi_2+3\eta_5\kappa_4+2\eta_8\kappa_7)x_1x_2+\ldots.
\end{align*}
We conclude $\eta_8=\kappa_7=u_3=0$, $\xi_2=-\eta_5\kappa_4$. Further we have
\begin{align*}
E_1-2\eta_5\kappa_4x_3C &=3\kappa_4(\eta_0+\eta_5u_1)x_1x_3+3\eta_5(\kappa_0+\kappa_4u_2)x_2x_3+\ldots,\\
E_2-2\kappa_4\xi_2x_1C &=-\eta_5\kappa_4^2u_1x_1^2+3\kappa_4\xi_0x_1x_3-6\eta_5(\kappa_4^2-1)x_2x_3+\ldots,\\
E_3-2\eta_5\xi_2x_2C &=-\eta_5^2\kappa_4u_2x_2^2+3\eta_5\xi_0x_2x_3-6\kappa_4(\eta_5^2-1)x_1x_3+\ldots.
\end{align*}
Therefore $u_1=u_2=\eta_0=\kappa_0=\xi_0=0$ and $\eta_5=\pm1$, $\kappa_4=\pm1$.
By symmetry \refpart{S5} we may assume $\eta_5=\kappa_4=-1$, then $\xi_2=-1$ as well. 
The remaining coefficients give $u_0=-4$ and $v_1=v_2=v_3=0$.
We get the third transformation 
$(x_1,x_2,x_3)\mapsto(-x_1-x_2x_3,-x_2-x_1x_3,-x_3-x_1x_2)$
of Theorem \ref{th:class}, on the cubic surface $x_1^2+x_2^2+x_3^2+x_1x_2x_3=4$. This concludes the proof of  Theorem \ref{th:class}.

%

\section{Kitaev's quadratic transformation on the monodromy manifold}\label{se:kit}

In this section we prove Theorem \ref{th:kit}. Actually, we prove that Kitaev's quadratic transformation (\ref{eq:kit1}) acts on 
${\bf a}$ as follows:
$$
{\bf a}=(0,a_t,a_1,0,a_{0t},a_{01},a_{t1})\to\tilde{\bf a}=(a_1,a_t,a_1,a_t,a_t a_1-a_{0t}a_{01}-a_{t1},2-a_{01}^2,a_{t1}).
$$
This is equivalent to Theorem \ref{th:kit} thanks to Okamoto birational transformations recalled in Appendix \ref{se:bir}. In fact$$ \textstyle
s_\rho s_0 s_1 s_t s_\rho s_t r_t\left(\frac{1}{2},\theta_t,\theta_1,\frac{1}{2}\right)=
(\widetilde\theta_1,\widetilde\theta_t,\widetilde\theta_t,\widetilde\theta_1+1),
$$
and 
$$
s_2 s_4 s_3 s_2 s_1 s_0 r_1 s_\rho s_\infty(\theta_1,\theta_t,\theta_1,\theta_t) = 
(0,2\widetilde\theta_t,0,2\widetilde\theta_1+1),
$$
where 
$$
\widetilde\theta_1= 1-\frac{\theta_t}{2}-\frac{\theta_1}{2},\qquad
\widetilde\theta_t= \frac{1}{2}-\frac{\theta_t}{2}+\frac{\theta_1}{2}.
$$

Observe that three of the quadratic transformations, namely Manin's, R.G.T. and T.O.S.,  act rather nicely on $(q,t)$. However, explicit expression for
Kitaev's transformation
involves either the derivative of $q(t)$ or the conjugate momentum $p(t)$.
In \cite{vidkit}, variations of Kitaev's transformations are formulated in terms
of an Okamoto transformation of $q(t)$. Here is a formulation in the same vein.
\begin{prop}\label{th:vkit}
Suppose that $Y_0(T)$ is a solution of $PVI(1/2,b,a,1/2)$.
Let us denote $Y_1(T)=s_{\rho}s_\infty s_1s_t Y_0(T)$,
which is a solution $PVI\left(\frac{a+b+1}2, \frac{a-b}2, \frac{a-b}2,
\frac{a+b+3}2\right)$.
Then $y_0(t)$ is a solution of $PVI (a, b, a, b)$, where
\begin{equation*}
y_0=\frac{(Y_1+\sqrt{T})\,\left((a-b)Y_0Y_1-(a+b)\sqrt{T}\,Y_0+2a\sqrt{T}\,Y_1\right)}
{4\sqrt{T}\,Y_1\,(aY_1-bY_0)}, \qquad
t=\frac{(\sqrt{T}+1)^2}{4\sqrt{T}}.
\end{equation*}
\end{prop}
\proof This is the inverse statement of \cite[Theorem 2.3]{vidkit}.
In the notation of that theorem, $Y_1=K_{[1/2,-a,-b,3/2]} Y_0$.
Note that the argument order $P_{VI}(\theta_0,\theta_1,\theta_t,\theta_\infty)$
rather than $PVI(\theta_0,\theta_t,\theta_1,\theta_\infty)$ is used in
\cite{vidkit}.
\endproof

Despite the complicated nature of Kitaev' quadratic transformation, its huge merit is that it is realised on the Fuchsian system as the composition of a rational transformation of the auxiliary variable $\lambda$ and a gauge transformation \cite{kitaev}. Our proof heavily relies on this construction which we resume here, omitting all the details.

We start from the initial Fuchsian system in the variable $\lambda$ with monodromy matrices $M_0,M_t,M_1$ with respect to the basis of loops $\Gamma_0, \Gamma_t, \Gamma_1,
\Gamma_\infty$ satisfying the following ordering (see Figure 1):
\be\label{eq:ord}
\Gamma_1\Gamma_t\Gamma_0=\Gamma_\infty^{-1}.
\ee

We then perform a rational transformation of the auxiliary variable $\lambda$:
\be\label{eq:rat}
\lambda=\mu^2,
\ee
so that the one obtains a new Fuchsian system in the form:
\be\label{eq:new-fuchs}
\frac{{\rm d}\Phi}{{\rm d}\mu} =\left(\frac{2A_0}{\mu}+\frac{A_t}{\mu-\tau}+\frac{A_t}{\mu+\tau}+
\frac{A_1}{\mu-1}+\frac{A_1}{\mu+1}
\right)\Phi,
\ee
where $\tau^2=t$. Kitaev proves that now $0$ and $\infty$ are apparent singularities and eliminates them by a rational gauge transformation leading to the intermediate Fuchsian system with $4$ poles, $\pm \tau,\pm 1$. Then he performs a conformal transformation
$$
\nu=\frac{(\mu+1)(\tau+1)}{2(\mu+\tau)}, 
$$
mapping
$$
-1\to 0, \quad 1\to 1,\quad -\tau\to\infty,\quad \tau\to T:= \frac{(\tau+1)^2}{4\tau},
$$
and leading to the final Fuchsian system with $4$ poles, $0,T,1,\infty$:
\begin{equation}\label{eq:fuchs1}
\frac{{\rm d}\tilde \Phi}{{\rm d} \nu} = \left(\frac{\tilde A_0}{\nu}+\frac{\tilde A_t}{\nu-T}+
\frac{\tilde A_1}{\nu-1}
\right)\tilde\Phi.
\end{equation}
Finally Kitaev proves that correspondingly the solutions $q(t)$ of the sixth Painlev\'e undergo a quadratic transformation (see formula (23) in \cite{kitaev}).

Our aim is to produce the corresponding transformation on the monodromy matrices, i.e. to express the monodromy matrices of the final Fuchsian system in terms of the initial ones. 

Let us concentrate on the first step: the rational transformation of the auxiliary parameter $\lambda$. Let us choose a basis of loops in  $\PP_\mu$, according to the following ordering (see Figure 2):
$$
\gamma_{1}^\mu\gamma_{\tau}^\mu\gamma_{0}^\mu\gamma_{-1}^\mu\gamma_{-\tau}^\mu
=\left(\gamma_{\infty}^\mu\right)^{-1}.
$$

\begin{figure}
\[ 
\begin{picture}(310,530)
\put(0,270){\resizebox{260pt}{!}{\includegraphics{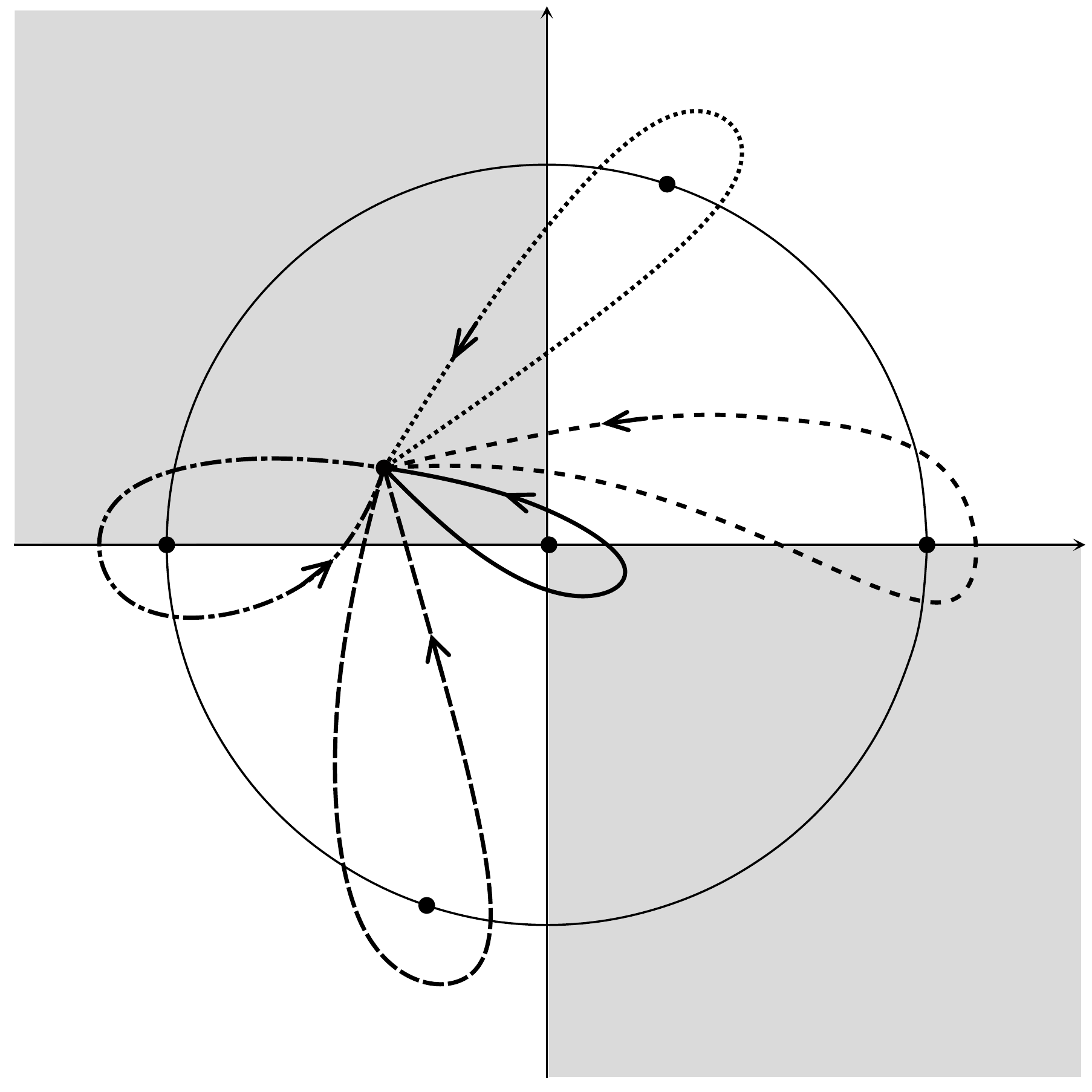}}}
\put(0,0){\resizebox{260pt}{!}{\includegraphics{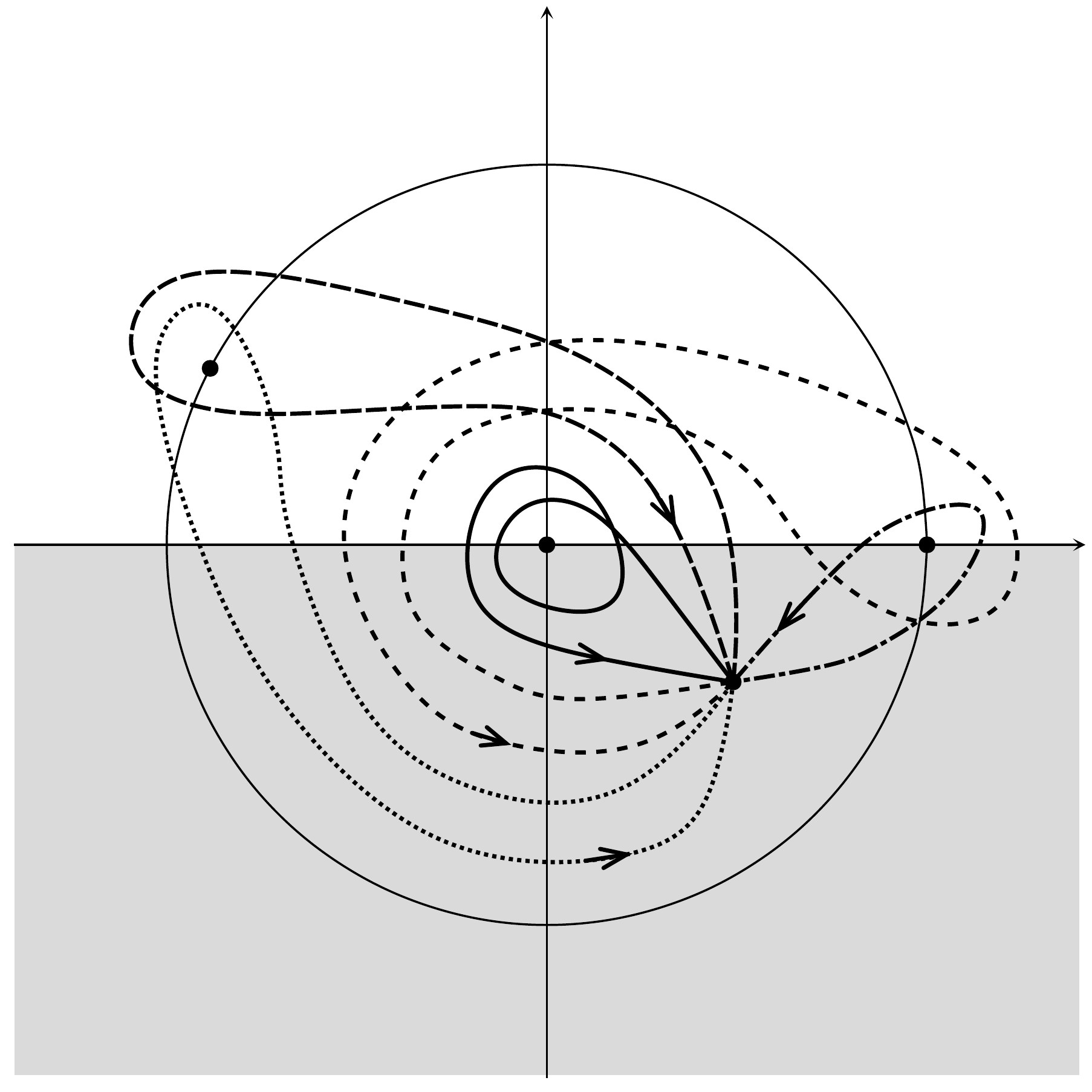}}}
\put(133,392){0} \put(206,403){$-1$} \put(43,403){$1$}
\put(141,474){$-\sqrt{t\,}$} \put(93,320){$\sqrt{t\,}$}
\put(147,382){$\gamma^\mu_0$}  \put(231,415){$\gamma^\mu_{-1}$}
\put(14,381){$\gamma^\mu_{1}$} \put(178,498){$\gamma^\mu_{\!-\!\sqrt{t\,}}$} 
\put(76,294){$\gamma^\mu_{\sqrt{t\,}}$} 
\put(280,310){\vector(0,-1){90}}  \put(288,261){$\lambda(\mu)=\mu^2$}
\put(276,318){$\PP^1_\mu$} \put(276,206){$\PP^1_\lambda$}
\put(133,122){0} \put(214,122){1} \put(43,174){$t$}
\put(147,110){$\gamma^\lambda_0$}  \put(181,177){$\gamma^\lambda_{-1}$}
\put(190,91){$\gamma^\lambda_{1}$} \put(84,57){$\gamma^\lambda_{\!-\!\sqrt{t\,}}$} 
\put(94,195){$\gamma^\lambda_{\sqrt{t\,}}$} 
\end{picture}
\] 
\caption{Path transformation under the quadratic covering}
\label{fig3}
\end{figure}
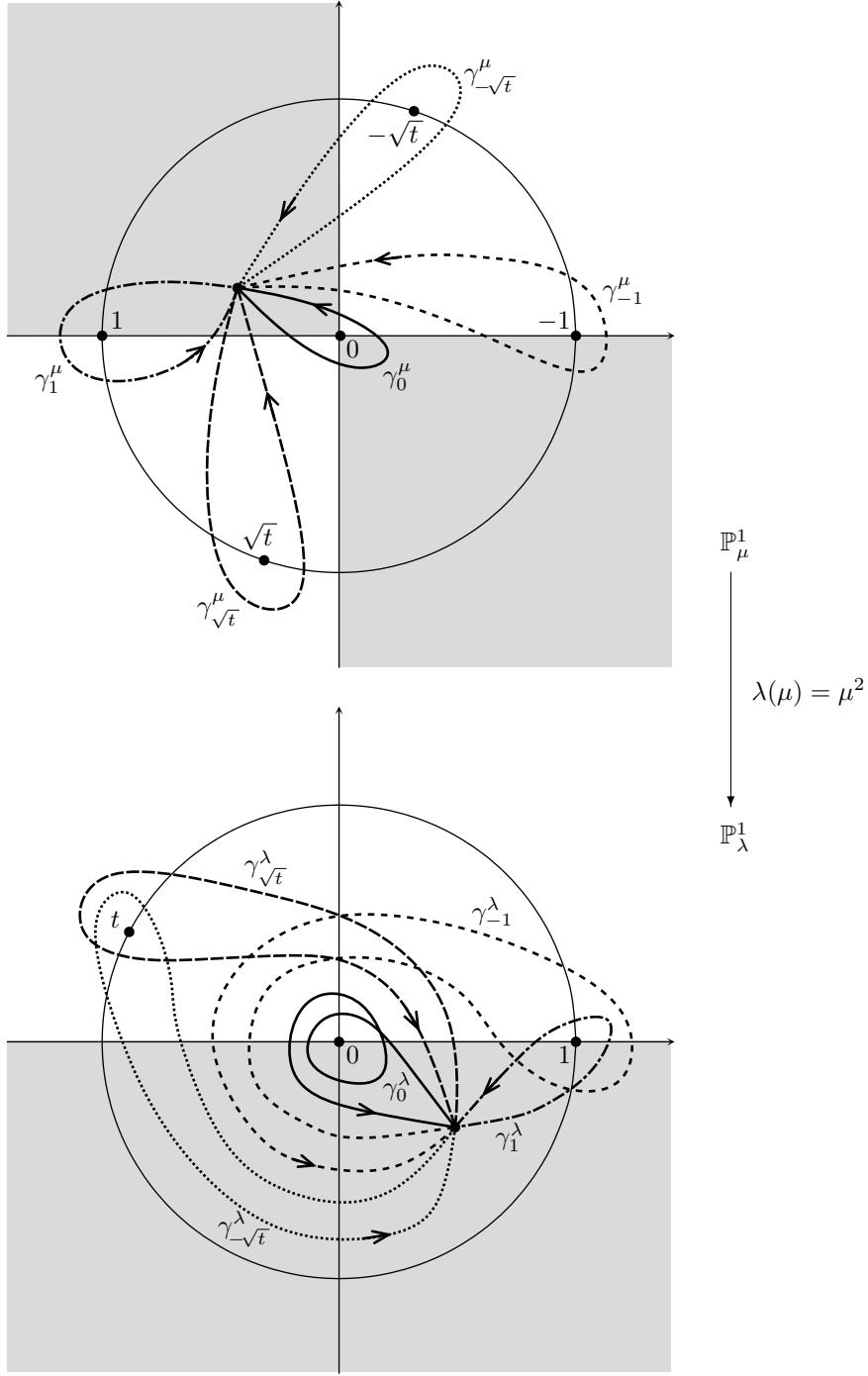


To draw the loops we use the fact that the preimage of the upper half-plane of $\PP^1_\lambda$
consists of the first and third quadrants in $\PP^1_\mu$. Following the four basic 
$\mu$-paths through the quadrants allows us to draw their projections in $\PP^1_\lambda$ easily. 
The images $\gamma^\lambda_i$, $i=\pm 1,\pm \tau,0,\infty$ in $\PP_\lambda$, of the basic loops  $\gamma^\mu_i$, $i=\pm 1,\pm \tau,0,\infty$, in  $\PP_\mu$ under the double--covering (\ref{eq:rat}) are:
 \begin{eqnarray}
 \label{eq:images}
&&
\gamma_{1}^\lambda=\Gamma_1,\quad 
\gamma_\tau^\lambda=\Gamma_t^,
\quad \gamma_{0}^\lambda=\Gamma_0^{2},\\
&&
\gamma_{-1}^\lambda=\Gamma_0^{-1}\Gamma_1 \Gamma_0,
\quad\gamma_{-\tau}^\lambda=\Gamma_0^{-1}\Gamma_t\Gamma_0,
\quad\gamma_\infty^\lambda=\Gamma_\infty^{2}.\nn
\end{eqnarray}

Note that the ordering of the two bases of loops are compatible, i.e. the images satisfy the relation:
$$
\gamma_1^\lambda\gamma_{\tau}^\lambda\gamma_{0}^\lambda\gamma_{-1}^\lambda
\gamma_{-\tau}^\lambda=\left(\gamma_{\infty}^\lambda\right)^{-1},
$$
provided that the basic loops  $\Gamma_0, \Gamma_t, \Gamma_1,
\Gamma_\infty$ satisfy (\ref{eq:ord}).

Observe that since the rational gauge transformations do not affect the monodromy matrices, the second step of Kitaev procedure will not play act on the monodromy matrices. The third step, i.e. the conformal transformation, only affects the labelling of the loops, or equivalently of the monodromy matrices, so we  can now deduce the final transformation on the monodromy matrices:
\be\label{eq:tr-mon}
\tilde M_0= M_0M_1 M_0^{-1},\qquad \tilde M_T= M_t,
\qquad \tilde M_1=M_1,\qquad \tilde M_\infty=M_0 M_t M_0^{-1}.
\ee
Note that $\tilde M_0\tilde M_t\tilde M_1\tilde M_\infty=\ID $ since $M_0^2=M_\infty^2=-\ID$.

By using (\ref{eq:tr-mon}) it is straightforward to obtain the following transformation on the monodromy manifold:
\be\label{eq:tr-pij}
\tilde a_0=a_1,\qquad \tilde a_t=a_t,\quad \tilde a_1=a_1,\quad
\tilde a_{0t}=a_ta_1-a_{0t}a_{01}-a_{t1} , \quad \tilde a_{01} = 2-a_{01}^2,\quad
\tilde a_{t1}= a_{t1}.
\ee
By using Okamoto birational transformation, we conclude the proof of Theorem \ref{th:kit}. \hfill{$\triangle$}\vskip 3mm

\section{Quartic transformation on the monodromy manifold.}\label{se:q4}

In this section we prove Theorem \ref{th:q4}.  To simplify the computations, we deal with three different quartic transformations according to the following diagram:
\[
\begin{picture}(282,148)(0,-65)
\put(0,60){$P_{VI}(\vartheta,\vartheta,\vartheta,\vartheta+1)$} 
\put(186,60){$P_{VI}(0,0,0,2\thinf)$}
\put(5,0){$P_{VI}\left(\frac12,\frac12,\frac12,\thinf\right)$}
\put(178,0){$P_{VI}(\thinf,\thinf,\thinf,\thinf)$}
\put(148,-60){$P_{VI}(1\!-\!\thinf,1\!-\!\thinf,1\!-\!\thinf,1\!-\!\thinf)$}
\put(120,69){$\Psi_{VI}^{[4]}$}  \put(126,9){$\tilde \zeta$}
\put(45,14){\vector(0,1){40}} \put(20,34){$s_\rho s_\infty$}
\put(225,14){\vector(0,1){40}}  \put(225,54){\vector(0,-1){40}} 
\put(186,34){$s_\infty s_\rho s_\infty$}
\put(225,-46){\vector(0,1){40}}  \put(225,-6){\vector(0,-1){40}} 
\put(211,-26){$s_\rho$}
\put(84,3){\vector(1,0){90}}  \put(84,63){\vector(1,0){90}}
\put(68,-10){\vector(3,-1){102}} \put(126,-24){$\zeta$}
\end{picture}
\]
where $\vartheta=\frac{\theta_\infty}{2}-\frac{1}{4}$.

As shown in the diagram, these three transformations are all related by Okamoto birational transformations, each of them is "simpler" for a specific task:  $\psi_{VI}^{[4]}$ is the one which transforms the solutions of PVI most neatly, $\zeta$ is the one which is directly obtained by composing two Kitaev's transformations up to symmetries (we show in the next page that $\zeta= 
\sigma_{1\infty}\cdot\hbox{Kitaev}\cdot\sigma_{0t}\cdot\hbox{Kitaev}$), and $\widetilde\zeta$ will be the one which we build by a single pull-back transformations of the associated Fuchsian system (see end of this Section). Since according to \cite{IIS} the transformations $s_\infty$ and $s_\rho$ act as identity on the monodromy manifold, we can deduce that these three transformations act on the monodromy manifold in the same way. 

\begin{remark}
Note that the most direct transformation obtained composing two Kitaev's quadratic transformations without the use of Okamoto symmetries is given by $P_{VI}\left(\frac12,1-\thinf,\frac12,\frac12\right)\to P_{VI}(1\!-\!\thinf,1\!-\!\thinf,1\!-\!\thinf,1\!-\!\thinf)$. However, this transformation requires a  renormalization of the target Fuchsian system.
\end{remark}

Let us recall the formulae for  the transformation $\psi^{(4)}_{VI}$ \cite{OST}:
$$
 \psi^{(4)}_{VI}(p,q,t)= \left(\tilde p,\tilde q,t\right),
 $$
with 
\begin{eqnarray*}
\tilde q \equal \frac{(q^2-t)^2}{4q(q-1)(q-t)},\\
\tilde p \equal \frac{4q(q-1)(q-t)\left[4q(q-1)(q-t)p-(4\thinf+\frac{1}{2})(q(q-1)+q(q-t)+(q-1)(q-t))\right]}
{(q^2-t)(q^2-2q+t)(q^2-2qt+t)}.
\end{eqnarray*}
In order to keep track of these transformations, we use the following notation:
 \begin{eqnarray*}
(q,p)\qquad\qquad\qquad & - & 
\mbox{a solution of $P_{VI}\left(\vartheta,\vartheta,\vartheta,\vartheta \right)$ 
for $\vartheta=\frac{\theta_\infty}{2}-\frac{1}{4}$},\\
(\tilde q,\tilde p)= \psi^{(4)}_{VI}(p,q,t) \quad& - & 
\mbox{a solution of $P_{VI}\left(0,0,0,4\vartheta-1\right)$ for $\vartheta=\frac{\theta_\infty}{2}-\frac{1}{4}$},\\
(y,p)=s_\infty s_\rho (q,p) \hspace{13.5pt} & - & 
\mbox{a solution of $P_{VI}\left(\frac12,\frac12,\frac12,\thinf\right)$},\\
\left(\tilde y, \tilde p\right)=s_\infty s_\rho (\tilde q, \tilde p) \hspace{13.5pt} & - & 
\mbox{a solution of } P_{VI}\left(1-\thinf,1-\thinf,1-\thinf,1-\thinf\right),\\
\left(\hat y, \tilde p\right)=s_\infty s_\rho s_\infty (\tilde q, \tilde p) & - & 
\mbox{a solution of } P_{VI}\left(\thinf,\thinf,\thinf,\thinf\right).
\end{eqnarray*}
Then we have 
\begin{eqnarray*}
y=q+\left(\frac34-\frac{\thinf}2 \right)\frac{1}{p},\qquad
\tilde y=\tilde q+\frac{ 1-\thinf }{\tilde{p}}, \qquad
\hat y=\tilde q+\frac{ \thinf }{\tilde{p}}. 
\end{eqnarray*}

Let us now see how to construct $\zeta(y,p)=(\tilde y, \tilde p)$. Let us fix the parameters of the Painlev\'e sixth equation in such a way that we can apply Kitaev's quadratic transformation twice (up to birational canonical transformations): $\theta_0=\theta_t=\theta_1=\frac{1}{2}$ and keep 
$\theta_\infty$ arbitrary. This in particular means that
$$
M_0^2=M_t^2=M_1^2=-\ID,\quad\hbox{so that}\quad M_\infty=-M_1 M_t M_0.
$$
On the level of the monodromy matrices we proceed as follows:
$$ \!\!
\begin{array}{ll}
(M_0,M_t,M_1,M_\infty) \hspace{-8pt}
& \flarrow{\sigma_{1\infty}}
 (M_0,M_t,-M_\infty,-M_\infty^{-1} M_1 M_\infty)   \\
& \flarrow{\mbox{\rm\tiny Kitaev}}
 (-M_0 M_\infty M_0^{-1},M_t ,- M_\infty,M_0 M_t M_0^{-1} )\\
& \flarrow{\sigma_{0t}}
(M_t,-M_t^{-1} M_0 M_\infty M_0^{-1} M_t,- M_\infty, M_0 M_t M_0^{-1})\\
& \flarrow{\mbox{\rm\tiny Kitaev}}
 (-M_t M_\infty M_t^{-1},-M_t^{-1}M_0 M_\infty M_0^{-1} M_t, -M_\infty,-M_0 M_\infty M_0^{-1})
 \qquad \\
 & \qquad\quad =(-M_t M_\infty M_t^{-1},M_tM_0 M_1,M_1M_tM_0,M_0 M_1 M_t).
\end{array}
$$
This corresponds to 
\begin{eqnarray}\label{eq:qtrm}
&&
(a_0,a_t,a_1,a_\infty,a_{0t},a_{01},a_{t1})=(0,0,0,a_\infty,a_{0t},a_{t1},a_{10})
\to\nn\\
&&\nn
\qquad \to (-a_\infty,-a_\infty,-a_\infty,-a_\infty,2-a_{t1}^2,2-a_{01}^2,2-a_{0t}^2),
\end{eqnarray}
which leads to (\ref{eq:qm}) by Okamoto birational transformations.

To show that this is the desired transformation, we need to prove that $\zeta$ acts on $(p,q)$ as
$s_\infty s_\rho s_\infty \psi^{(4)}_{VI} s_\rho s_\infty$, i.e. we need to prove the following formulae in terms of $(y,p)$:
\begin{eqnarray}  \label{eq:trn}
\tilde y = \frac{L_1\,L_2}{4\,p\,L_5},
\qquad   \tilde p = \frac{16p\,L_3\,L_5}{L_1\,L_4},
\end{eqnarray}
where $L_1=\left(py+\frac{2\thinf-3}4\right)^2-tp^2,\quad$  
$L_2 = \textstyle \left(py+\frac{2\thinf-3}4\right) \left(py-\frac{2\thinf-1}4\right)-tp^2$,
\begin{eqnarray*} 
L_3 \equal \textstyle \left(py+\frac{2\thinf-3}4\right)
\left(py-p+\frac{2\thinf-3}4\right)\left(py-tp+\frac{2\thinf-3}4\right), \\
L_4 \equal \textstyle \left(\left(py-p+\frac{2\thinf-3}4\right)^2+(t-1)p^2\right)
\left(\left(py-tp+\frac{2\thinf-3}4\right)^2-t(t-1)p^2\right), \\
L_5 \equal \textstyle \left(py-\frac{\thinf}4\right) \! \left(py+\frac{2\thinf-3}4\right)^2 \!
-p\,(t+1) \! \left(py-\frac14\right) \! \left(py+\frac{2\thinf-3}4\right)+tp^2\left(py+\frac{\thinf-2}4\right).
\end{eqnarray*}
In order to prove the same formulae from pullback transformation,
we are going to build the quartic transformation $\widetilde\zeta$ on the PVI directly as a rational-pull-back transformations of the corresponding Fuchsian system, i.e. by a unique RS-transformation rather than the composition of two of them. The transformation rule $\widetilde\zeta(y,p)=(\hat y,\tilde p)$ is rather less pretty than (\ref{eq:trn}):
$$
\hat y =\frac{L_1\,L_6}{4L_3\,L_5}, 
$$
where 
\begin{eqnarray*} 
L_6 \equal \textstyle y \! \left(py+\frac{2\thinf-3}4\right)^{4}
\! -(t+1) \! \left(py+\frac{2\thinf-1}4\right) \! \left(pq+\frac{2\thinf-3}4\right)^{3} \!+ 
\frac{4\thinf-1}2tp \! \left(py+\frac{2\thinf-3}4\right)^{2}+ \\
\cequal \textstyle 
+t(t+1)p^2\left(py-\frac{2\thinf+1}4\right)\left(py+\frac{2\thinf-3}4\right)-p^3t^2\left(py-\frac12\right).\end{eqnarray*}
The transformation $\widetilde\zeta$ is realised as the composition of a rational transformation $R$ of the auxiliary variable $\lambda$ and a gauge transformation $S$.  

The rational transformation is given by:
$$
\lambda=R(\mu)=\frac{(\mu^2-t)^2}{4\mu(\mu-1)(\mu-t)},
$$
note that $R$ has the same form as the folding transformation $\psi_{VI}^{[4]}$ on $q$.
This maps the initial Fuchsian system to a new Fuchsian system
 \begin{eqnarray}
& \displaystyle \frac{{\rm d}\Phi}{{\rm d}\mu}=
&\left(\frac{A_\infty}{\mu}+\frac{A_\infty}{\mu-t}+\frac{A_\infty}{\mu-1}+
 \frac{2 A_0}{\mu-e_1}+ \frac{2 A_0}{\mu-e_2}+ \right.\label{eq:fumu}\\
 &&
\left.+ \frac{2 A_t}{\mu-e_3}+ \frac{2 A_t}{\mu-e_4}+
  \frac{2 A_1}{\mu-e_5}+ \frac{2 A_1}{\mu-e_6}
 \right)\Phi,\nn\end{eqnarray}
where $e_1,\dots,e_6$ are the roots of the following quadratic equations
$$
\begin{array}{lc}
e_i^2-t=0, & i=1,2,\\
e_i^2-2 s_i+t=0, & i=3,4,\\ 
e_i^2-2 t s_i+t=0& i=5,6.
\end{array}
$$
It is worth observing that the above quadratic equations emerge as the numerators of $R$, $R-1$ and $R-t$ respectively:
$$
R(\mu)-1=\frac{(\mu^2-2\mu+t)^2}{4\mu(\mu-1)(\mu-t)},\qquad
R(\mu)-t=\frac{(\mu^2-2t\mu+t)^2}{4\mu(\mu-1)(\mu-t)}.
$$

We now construct the gauge transformation by imposing that the apparent singularities $e_1,\dots,e_6$ have to be removed. This gauge transformation must have the form
\begin{eqnarray*}
\frac1{\sqrt{(\mu^2-t)(\mu^2-2\mu+t)(\mu^2-2t\mu+t)}}
\left( \begin{array}{cc} G_{1,1} & G_{1,2} \\ G_{2,1} & G_{2,2} \end{array} \right),
\end{eqnarray*}
with $G_{1,1}$, $G_{1,2}$, $G_{2,1}$, $G_{2,2}$ polynomials in $\mu$ of degree $3,2,2,3$, respectively, because:
\begin{itemize}
\item the local exponents $1/2$ at the six apparent singularities must be shifted to $0$,
hence the denominator;
\item the transformation matrix must be asymptotically the identity as $\mu\to\infty$,
since we keep the local exponents at $\mu=\infty$ the same as in $\lambda=\infty$;
this gives the degree bounds.
\end{itemize}
Besides, the local exponents $-1/2$ at the six singular points must be shifted to 0 as well.

In order to carry out our computations it is better to parametrize the matrices $A_0,A_t,A_1$ as follows:
\begin{equation}
A_k(\lambda)=\frac12\, \left(\begin{array}{cc}
u_k & w_k\,(\theta_k-u_k) \vspace{1pt} \\
\displaystyle\frac{\theta_k+u_k}{w_k} & -u_k 
\end{array}\right) \quad \mbox{for } k\in\{0,1,t\},
\end{equation}
where
\begin{eqnarray} \label{eq:www}
\qquad w_0=\frac{k\,q}{t\,(u_0-\theta_0)},\quad
w_1=\frac{k\,(q-1)}{(1-t)\,(u_1-\theta_1)},\quad
w_t=\frac{k\,(q-t)}{t\,(t-1)\,(u_t-\theta_t)},  
\end{eqnarray}
and
\begin{eqnarray}\label{eq:uuu}
u_0 \equal \frac1{2\theta_{\infty}} \left(
\frac{s^2-2\theta_\infty t s-\theta_\infty^2 t q (q-t-1)}{t\,(q-1)\,(q-t)}-\theta_0^2
+\frac{\theta_1^2(t-1)q}{t(q-1)}-\frac{\theta_t^2(t-1)q}{q-t} \right),\nn\\
u_1 \equal \frac1{2\theta_{\infty}} \left(
\frac{s^2-2\theta_\infty q s+\theta_\infty^2 t q (q-t+1)}{(1-t)\,q\,(q-t)}-\theta_1^2
+\frac{\theta_0^2t(q-1)}{(t-1)q}+\frac{\theta_t^2t(q-1)}{q-t} \right),\nn\\
u_t \equal \frac1{2\theta_{\infty}} \left(
\frac{s^2-2\theta_\infty t q s+\theta_\infty^2 t q (q+t-1)}{t\,(t-1)\,q\,(q-1)}-\theta_t^2
+\frac{\theta_0^2(q-t)}{(t-1)q}+\frac{\theta_1^2(q-t)}{t(q-1)} \right),\nn\\
\end{eqnarray}
where
\begin{eqnarray*}
s=\theta_0(q-1)(q-t)+\theta_1q(q-t)+(\theta_t-\theta_\infty)q(q-1)+\theta_\infty t q-2q(q-1)(q-t)p.
\end{eqnarray*}
Here we have replaced $p$ by the parameter $s$ which gives an attractive parametrization of the particular traceless normalization of the $2\times2$ Fuchsian system because $s=t(q-1)u_0+(t-1)qu_1+\theta_\infty$. 

Note that in the above formulae for $A_0,A_t,A_1$, $q$ denotes the generic solutions of $PVI(\theta_0,\theta_t,\theta_1,\theta_\infty)$, so that for the initial Fuchsian system we need to replace $q$ by $y$ and $(\theta_0,\theta_t,\theta_1,\theta_\infty)$ by $\left(\frac{1}{2},\frac{1}{2},\frac{1}{2},\theta_\infty\right)$ and for the final Fuchsian system we need to replace $q$ by $\hat y$ and $(\theta_0,\theta_t,\theta_1,\theta_\infty)$ by $(\theta_\infty,\theta_\infty,\theta_\infty,\theta_\infty)$.

The  local solutions of the initial system (after $R$ transformation and before gauge $S$) are:
\begin{eqnarray*}
\frac1{\sqrt{\lambda}}\left( {w_0\choose1}+O(\lambda) \right)& & \mbox{at $\lambda=0$},\\
\frac1{\sqrt{\lambda-1}}\left( {w_1\choose1}+O(\lambda-1) \right)& & \mbox{at $\lambda=1$},\\
\frac1{\sqrt{\lambda-t}}\left( {w_t\choose1}+O(\lambda-t) \right)& & \mbox{at $\lambda=t$}.
\end{eqnarray*}
After a direct pull-back, the local solutions must be:
\begin{eqnarray*}
\frac1{\sqrt{\mu^2-t}}\left( {w_0\choose1}+O(\mu^2-t) \right)& & \mbox{at $\mu=\pm\sqrt{t}$, \quad etc.}
\end{eqnarray*}
To kill the local exponents $-1/2$, we must have
\begin{eqnarray*}
\left( \begin{array}{cc} G_{1,1} & G_{1,2} \\ G_{2,1} & G_{2,2} \end{array} \right) 
{w_0\choose1} & \mbox{divisible by} & \mu^2-t,\\
\left( \begin{array}{cc} G_{1,1} & G_{1,2} \\ G_{2,1} & G_{2,2} \end{array} \right)
{w_1\choose1}  & \mbox{divisible by} & \mu^2-2\mu+t,\\
\left( \begin{array}{cc} G_{1,1} & G_{1,2} \\ G_{2,1} & G_{2,2} \end{array} \right)
{w_t\choose1} & \mbox{divisible by} & \mu^2-2t\mu+t. 
\end{eqnarray*}
This gives exactly enough linear relations for the coefficients of
$G_{1,1}$, $G_{1,2}$, $G_{2,1}$, $G_{2,2}$ (as polynomials in $\mu$)
to determine the gauge matrix up to scalar multiples or rows. We obtain:
\begin{eqnarray*}
G_{1,1} \equal (w_0-w_1)(w_0-w_t)(w_1-w_t)\mu^3\\
\cequal +(w_t(w_0-w_1)(w_0+w_1-2w_t)t-w_1(w_0-w_t)(w_0+w_t-2w_1))\mu^2\\
\cequal -(w_0^2+w_0w_1+w_0w_t-3w_1w_t)(w_1-w_t)t\mu \\
\cequal -(w_t(w_0-w_1)^2t-w_1(w_0-w_t)^2)t,\\
G_{1,2} \equal  
(w_t(w_0\!-\!w_1)(w_0w_t\!+\!w_1w_t\!-\!2w_0w_1)t
-w_1(w_0\!-\!w_t)(w_0w_1\!+\!w_1w_t\!-\!2w_0w_t))\mu^2\\
\cequal +2tw_0(w_1-w_t)(w_0w_t+w_0w_1-2w_1w_t)\mu\\
\cequal +(w_t^2(w_0-w_1)^2t-w_1^2(w_0-w_t)^2)t,\\
G_{2,1} \equal ((w_0-w_1)(w_0+w_1-2w_t)t-(w_0-w_t)(w_0+w_t-2w_1))\mu^2\\
\cequal -2t(w_1-w_t)(2w_0-w_1-w_t)\mu-((w_0-w_1)^2t-(w_0-w_t)^2)t,\\
G_{2,2} \equal  (w_0-w_1)(w_0-w_t)(w_1-w_t)\mu^3\\
\cequal +((w_0\!-\!w_1)(w_0w_t\!+\!w_1w_t\!-\!2w_0w_1)t
-(w_0\!-\!w_t)(w_0w_1\!+\!w_1w_t\!-\!2w_0w_t))\mu^2\\
\cequal +t(w_1-w_t)(3w_0^2-w_0w_t-w_0w_1-w_1w_t)\mu \\
\cequal +t(w_t(w_0-w_1)^2t-w_1(w_0-w_t)^2).
\end{eqnarray*}%
If we substitute in the formulae the expressions of $w_0, w_1,w_t,u_0,u_1,u_t$ given by formulae
(\ref{eq:www}) and (\ref{eq:uuu}) the above expressions do not simplify. A check that this transformation actually gives rise to the desired transformation law on $(p,q)$ is a straightforward but rather heavy computation. We have made a maple worksheet available;
see {\sf http://www.math.kobe-u.ac.jp/\~{}vidunas/PainleveQuartic.mw}. 

We are now going to prove that the corresponding transformation on the monodromy manifold is given by formulae (\ref{eq:qm}). Again, it is only the first transformation $\lambda=R(\mu)$ which carries all the information because the gauge transformation does not affect the way monodromy matrices of the system (\ref{eq:fumu}) depend on the initial ones which are computed with respect to the basis of loops 
$\Gamma_0,\Gamma_t,\Gamma_1$ shown in Figure 1.

We use the same technique as in the proof of Theorem \ref{th:kit}: we fix a basis of loops 
$\gamma_\infty^\mu,\gamma_0^\mu, \gamma_t^\mu,\gamma_1^\mu, \gamma_{e_1}^\mu,
\dots,\gamma_{e_6}^\mu$ in $\PP_\mu$ such that (see Figure 3):
\begin{equation}\label{eq:loopmu}
\gamma_\infty^\mu \gamma_1^\mu\gamma_t^\mu \gamma_0^\mu  \gamma_{e_1}^\mu,
\dots,\gamma_{e_6}^\mu=1.
\end{equation}
We construct their images in the $\mathbb P_\lambda$ as in Figure 3 by  marking a great circle 
(or a line) through $\lambda=0,1,t$, and choosing $z=\infty$ to be inside the shaded half-plane.
We assume a base point to lie in the other half-plane. We mark the 6 branching points
$e_1,\ldots,e_6$ in the $\mu$--plane by $0^*$, $1^*$ or $t^*$, depending on their 
images $\lambda=0$, $\lambda=1$ or $\lambda=t$ (Note that we do not draw the loops around $e_1,\dots,e_6$ because those singularities are apparent).The pre-image of the circle segment in $\PP^1_\lambda$ between $1$ and $t$ 
must be a {\em dessin d'enfant} for the Belyi covering $R(\mu)$, which is (topologically) 
the circle in $\mu$-plane with the marked points $1^*$, $t^*$. After adding the pre-images $0^*$
we obtain the pre images of the (shaded and white) half-planes of $\PP^1_\lambda$.
Topologically, the pre-image of the circle in $\PP^1_\lambda$ is an octahedral graph.
The pre-images $\mu=0$, $\mu=1$, $\mu=t$, $\mu=\infty$ of $\lambda=\infty$ lie in the 4
different shaded regions on $\PP^1_\mu$. The images of the basis paths $\gamma_0^\mu$,
$\gamma_t^\mu$, $\gamma_1^\mu$, $\gamma_\infty^\mu$ are obtained by following
which segments between the $0^*$, $1^*$, $t^*$ points they cross.

\begin{figure}
\[ 
\begin{picture}(350,530)
\put(0,270){\resizebox{260pt}{!}{\includegraphics{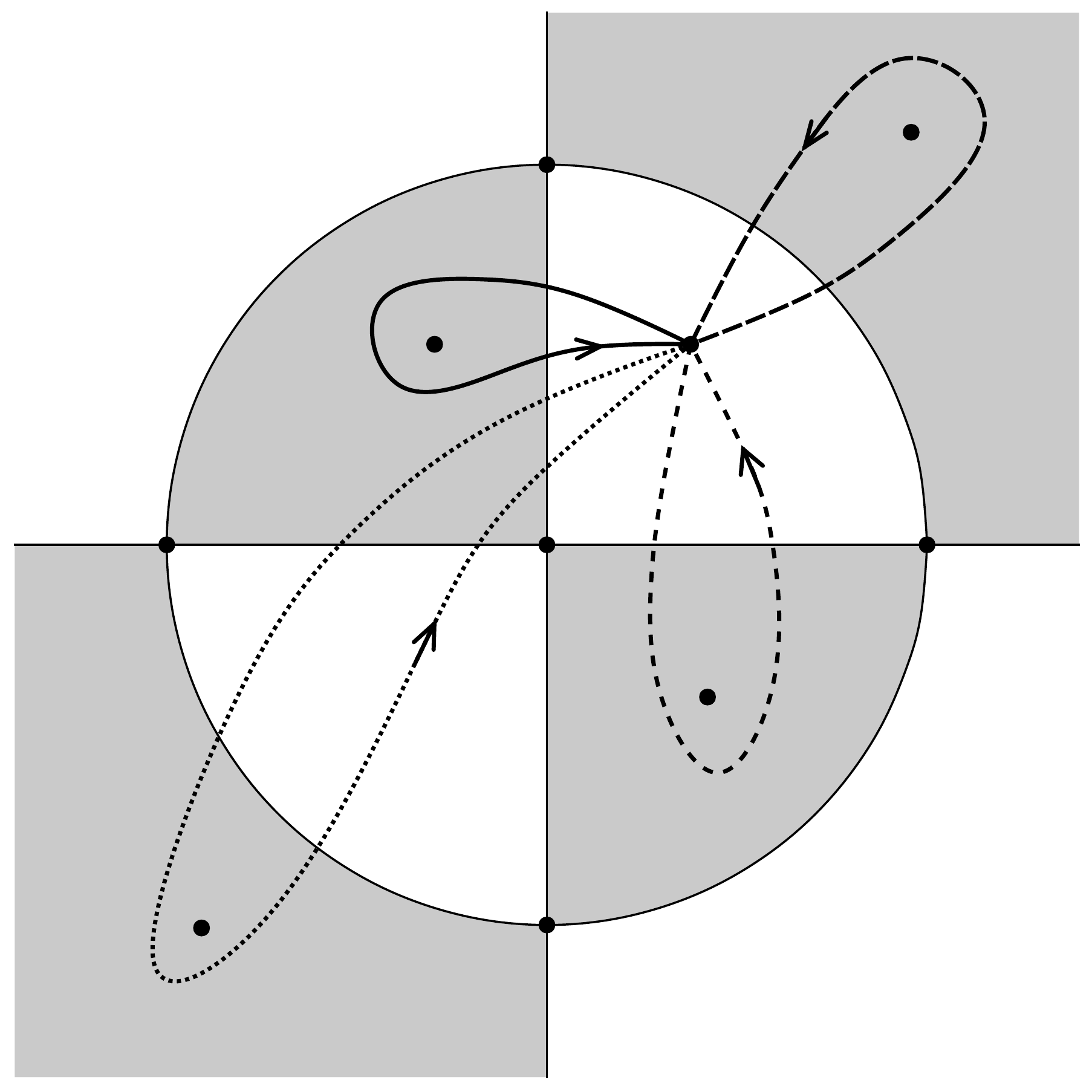}}}
\put(0,-20){\resizebox{280pt}{!}{\includegraphics{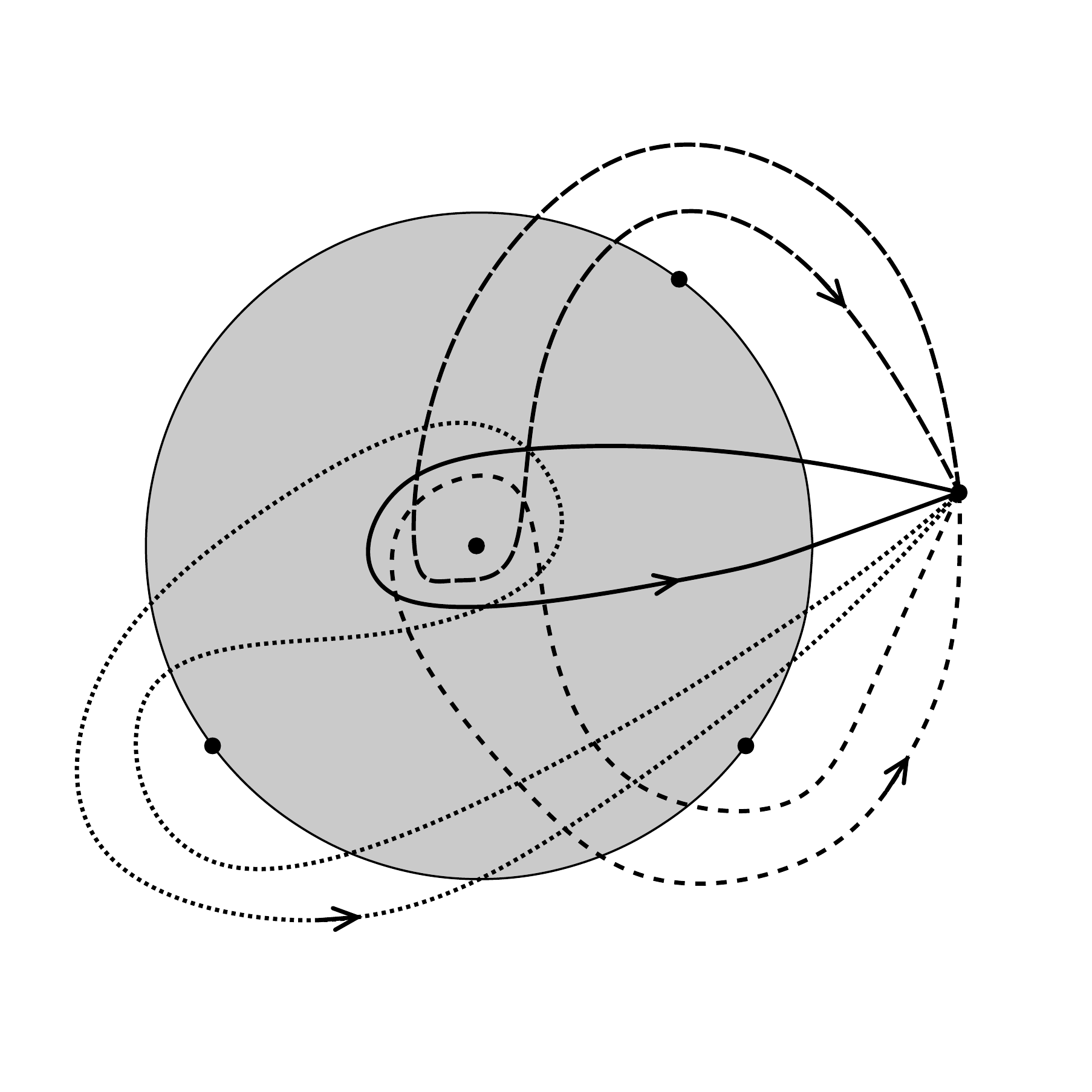}}}
\put(119,392){$0^\star$} \put(225,392){$t^\star$} \put(29,403){$t^\star$}
\put(133,301){$1^\star$} \put(120,494){$1^\star$} \put(104,452){$0$} 
\put(171,367){$1$} \put(209,496){$t$} \put(50,312){$\infty$}
\put(78,435){$\gamma^\mu_0$} \put(188,378){$\gamma^\mu_1$}  
\put(65,370){$\gamma^\mu_{\infty}$} \put(216,467){$\gamma^\mu_{t}$} 
\put(270,310){\vector(0,-1){90}}  \put(276,261){$\lambda(\mu)=\frac{(\mu^2-t)}{4\mu(\mu-1)(\mu-t)}$}
\put(266,318){$\PP^1_\mu$} \put(266,206){$\PP^1_\lambda$}
\put(197,66){0} \put(180,187){1} \put(46,66){$t$} \put(115,125){$\infty$}
\put(165,134){$\gamma^\lambda_0$}  \put(216,36){$\gamma^\lambda_{1}$} 
\put(14,97){$\gamma^\lambda_\infty$} \put(139,220){$\gamma^\lambda_{t}$} 
\end{picture}
\] 
\caption{Path transformation under the degree 4 covering}
\label{fig3}
\end{figure}
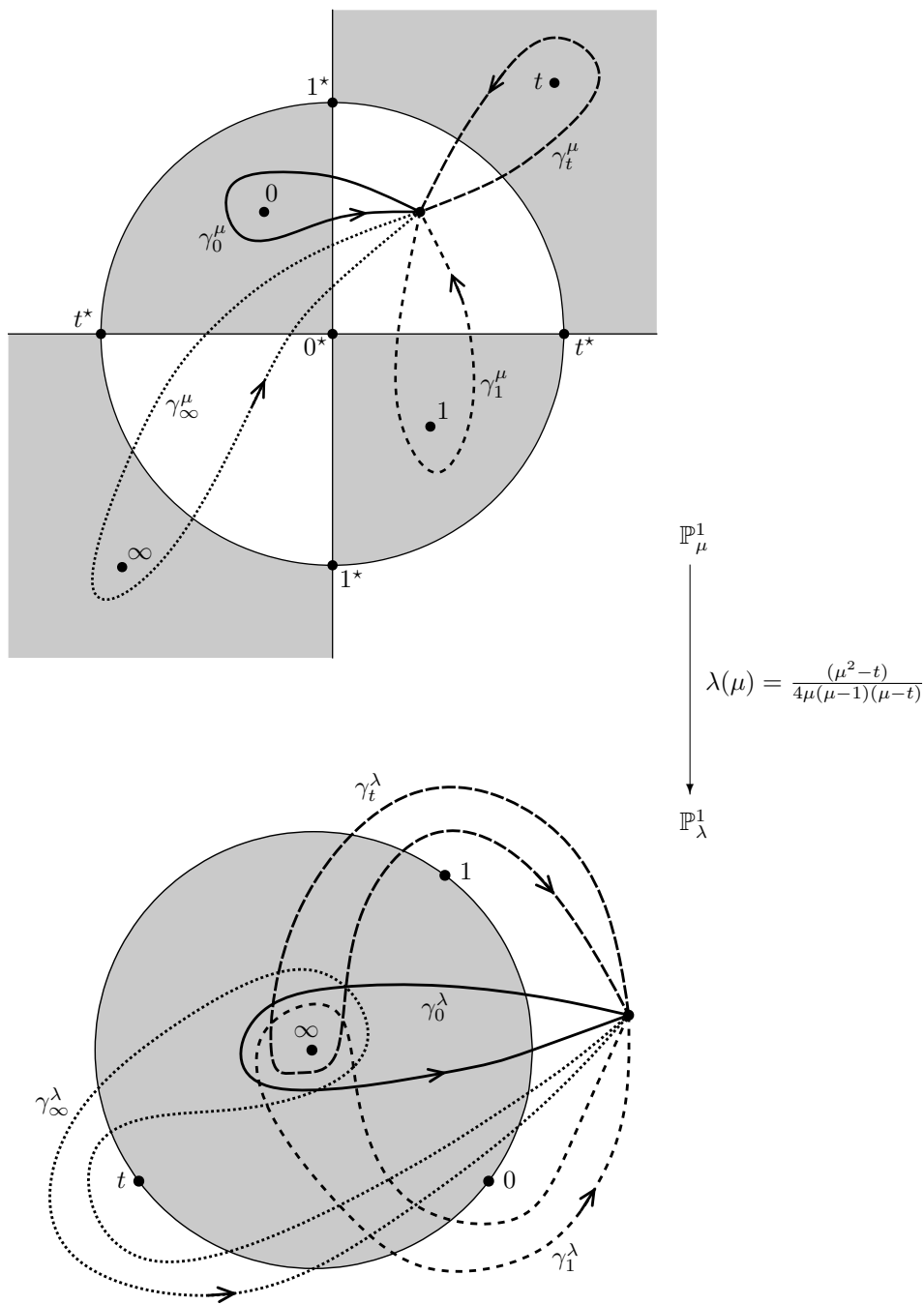


We see that:
$$
\gamma^\lambda_0=\Gamma_t^{-1}\Gamma_\infty\Gamma_t,\qquad
\gamma^\lambda_t=\Gamma_1^{-1}\Gamma_\infty\Gamma_1,\qquad
\gamma^\lambda_1=\Gamma_\infty,
$$ so that 
$$
\widetilde M_0=M_t^{-1}M_\infty M_t,\quad 
\widetilde M_t = M_1^{-1}M_\infty M_1,\quad
\widetilde M_1= M_\infty,
$$
from which we get (\ref{eq:qm}). \hfill{$\triangle$}.

\section{Picard case: proof of Theorem \ref{th:cubic}}\label{se:Pic}

Here we prove Theorem \ref{th:cubic}. We use the fact that the general solution of PVI in Picard case is given by
$$
q(t;\nu_1,\nu_2)=\wp \big(\nu_1 \omega_1(t) + \nu_2 \omega_2(t) ;
\omega_1(t) ,\omega_2(t) \big)  + {t+1\over3} 
$$
where $(\nu_1,\nu_2)\in\mathbb C^2$ are free parameters, and the half--periods 
$\omega_{1,2}(t)$ are two linearly independent solutions of the following hypergeometric equation:
$$
t(1-t)\omega''(t)+(1-2 t) \omega'(t)-{1\over4} \omega(t)=0.
$$
The free parameters $\nu_1,\nu_2$ are defined modulo $2$ and are generically (i.e. for $\nu_1,\nu_2\neq 0,1$) related to the monodromy data as follows \cite{M}:
\be\label{eq:vu-a}
a_{0t}= -2 \cos(\pi \nu_2),\qquad
a_{01}= -2 \cos(\pi (\nu_1-\nu_2)),\qquad
a_{t1}= -2 \cos(\pi \nu_1).
\ee
Let us consider the third transformation of Theorem \ref{th:class} on the monodromy manifold:
\be\label{eq:cumo}
(a_{0t},a_{01},a_{t1})\mapsto(-a_{0t}-a_{01}a_{t1},-a_{01}-a_{0t}a_{t1},-a_{t1}-a_{01}a_{0t}).
\ee
The corresponding transformation on PVI must map Picard solutions to Picard solutions:
$$
\wp (\nu_1 \omega_1(t) + \nu_2 \omega_2(t) ;\omega_1(t),\omega_2(t)) 
+ {t+1\over3}  \to
\wp (\tilde\nu_1 \omega_1(\tilde t) +
\tilde\nu_2 \omega_2(\tilde t);\omega_1(\tilde t),\omega_2(\tilde t)) 
+ {\tilde t+1\over3} ,
$$
where, by using  (\ref{eq:vu-a}) and  
(\ref{eq:cumo}):
$$
\left(\begin{array}{c}
\tilde\nu_1\\
\tilde \nu_2\\
\end{array}\right)=\left(\begin{array}{cc}
1&-2\\ 2&-1\\\end{array}\right)
\left(\begin{array}{c}
\nu_1\\ \nu_2\\
\end{array}\right),
$$
which leads to an isogeny of degree three on the elliptic curve
\be\label{eq:elliptic}
w^2=q(q-1)(q-t).
\ee

Producing generic isogenies of low degree is apparently a frequent routine
for those working on elliptic curves.
We computed a general form of a cubic isogeny between two elliptic curves 
in the Weiertstrass form using the V\'elu Theorem (see chapter 25 in \cite{Gal}).
Here are the elliptic curves and the isogeny.
\begin{align}\label{eq:isogen}
E_1: &\quad w^2=y^3+3a(a+2b)y+a(3b^2-a^2),\nn\\
E_2: &\quad W^2=Y^3-3a(19a+18b)Y-a(169a^2+252ab+81b^2),\\
&
Y=y+\frac{12a(a+b)(y+b)}{(y-a)^2},\quad W=w-\frac{12a(a+b)(y+a+2b)w}{(y-a)^3},\nn
\end{align}
The parameters $a,b$ are to be considered as a homogeneous pair.
The points with $y=a$ on $E_1$ are rational points of order 3.

To derive the transformation between solutions of the Painl\'eve sixth equation, we
translate this isogeny to a transformation between the elliptic curves in the Legendre form, i.e. between the elliptic curve  (\ref{eq:elliptic}) and
\be\label{eq:elliptic1}
\tilde w^2=\tilde q(\tilde q-1)(\tilde q-\tilde t).
\ee
As a first step to achieve this, we have to parameterise $y$ and $b$ in such a way that the cubic polynomial on the right end side of $E_1$ has a rational root $y_0$, then shift this rational root to $0$. We then repeat the procedure imposing a second rational root and shifting it to $1$. 

The parameters
$a,b$ form a homogeneous pair, so we can  put $a=1$ without loss of generality. To find the correct parameterisation, we impose that the discriminant in $b$ in equation
$$
y^3+3(1+2b)y+(3b^2-1)=0
$$
is a perfect square when evaluated at $y_0$. This leads to impose  $y_0=1-u^2/3$ and therefore  $b=u^3/9+u^2/3-1$. By shifting 
$y\mapsto y+1-u^2/3$ and  $Y\mapsto Y+1+4u+4u^2$
the two elliptic curves and the isogeny between them become:  
\begin{align*}
E'_1: &\quad w^2=y\left(y^2+(3-u^2)y+\frac{u^3(u+2)}3\right),\\
E'_2: &\quad W^2=Y\left(Y^2+3(1+4u+u^2)Y+3u(u+2)^3\right),\\
&Y=\frac{y\left(y-2u-u^2\right)^2}{(y-\frac13u^2)^2},\quad
W=w-\frac{4u^2(1+\frac13u)(y+\frac29u^3+\frac13u^2)}{(y-\frac13u^2)^3}\,w.
\end{align*}
Now we need to factorize the quadratic polynomial in $E'_1$, namely $y^2+(3-u^2)y+\frac{u^3(u+2)}3$.
The discriminant with respect to $y$ is equal to $-\frac13(u-1)(u+3)$,
and to make it a perfect square we substitute $u\mapsto3s/(s^2+s+1)$.
After the scalings\\
$y\mapsto -3(2s+1)y/(s^2+s+1)^2$, 
$Y\mapsto-3(2s+1)^3Y/(s^2+s+1)^2$\\
we get the new elliptic curves and the isogeny between them in the form:
\begin{align*}
&
E''_1: \quad w^2=-\frac{27(2s+1)^3}{(s^2+s+1)^6}\,y\,(y-1)\left(y-\frac{s^3(s+2)}{2s+1}\right),\\
&
E''_2: \quad W^2=-\frac{27(2s+1)^9}{(s^2+s+1)^6}\,Y(Y-1)\left(Y-\frac{s(s+2)^3}{(2s+1)^3}\right),\\
&
Y=\frac{y\left(y+s(s+2)\right)^2}{((2s+1)y+s^2)^2},
\, W=w-\frac{4s^2(s+1)^2((2s+1)(s^2+s+1)y-s^2(s^2+3s+1))w}{((2s+1)y+s^2)^3}.
\end{align*}
At the last step, we get rid of the front factors (in $s$) of the cubic polynomials by rescaling $w$.
Although the rescaling would require the square root $\sqrt{-3(2s+1)}$, the square roots
for $w$ and $W$ would cancel out, and we just have to divide $w$ (in both instances)
by $(2s+1)^3$ in the isogeny expression. The $y$-component does not change at all. 

And it is the $y$ component that gives the transformation of Painlev\'e solutions, namely we identify $y$ with $q$ and $Y$ with $\tilde q$ in the above formulae, thus proving  Theorem \ref{th:cubic}.

\setcounter{section}{0}

\def\thetheorem{A.\arabic{theorem}}
\def\theprop{A.\arabic{prop}}
\def\thelemma{A.\arabic{lm}}
\def\thecor{A.\arabic{cor}}
\def\theexam{A.\arabic{exam}}
\def\theremark{A.\arabic{remark}}
\def\theequation{A.\arabic{equation}}

\appendix{Isomonodromic deformations associated to the sixth Painlev\'e equation}\label{se:iso}

Here we recall without proof some very well known facts about the Painlev\'e sixth equation and its relation to the monodromy preserving deformations equations \cite{JMU,MJ1}.

The sixth Painlev\'e sixth equation (\ref{eq:P6})
describes  the monodromy preserving deformations of a rank $2$ meromorphic connection over 
$\PP^1$ with four simple poles $0,t,1$ and $\infty$:
\begin{equation}\label{eq:fuchs}
\frac{{\rm d} \Phi}{{\rm d} \lambda} = \left(\frac{A_0(t)}{\lambda}+\frac{A_t(t)}{\lambda-t}+\frac{A_1(t)}{\lambda-1}
\right)\Phi,
\end{equation}
where 
\begin{eqnarray}
&&{\rm eigen}(A_i)= \pm\frac{\theta_i}{2}, \quad\hbox{for } i=0,t,1,
\quad A_\infty:=-A_0-A_t-A_1\label{eq:eigen1}\\
&& 
A_\infty =
\left(\begin{array}{cc}\frac{\theta_\infty}{2}&0\\
0&-\frac{\theta_\infty}{2}\\
\end{array}\right),\label{eq:eigen2}
\end{eqnarray}
and the parameters $\theta_i$, $i=0,t,1,\infty$ are related to the PVI parameters by
(\ref{eq:param}).
The precise dependence of the matrices $A_0,A_t,A_1$ on the PVI solution $q(t)$ and its first derivative $\dot q(t)$ can be found in \cite{MJ1}, we will use a slightly modified parametrisation in Section \ref{se:q4}  which simplifies our formulae. In this paper we assume $\theta_\infty\not\in\mathbb Z$.

The solution $\Phi(\lambda)$ of the system (\ref{eq:fuchs}) is a multi-valued analytic function 
in the punctured Riemann sphere $\mathbb P^1\setminus\{0,t,1,\infty\}$ and its multivaluedness is 
described by the so-called monodromy matrices, i.e. the images of the generators of the fundamental group under the anti-homomorphism
$$
\rho:\pi_1\left( \mathbb P^1\backslash\{0,t,1,\infty\},\lambda_0\right)\to SL_2(\mathbb C).
$$
In this paper we fix the base point $\lambda_0$ at infinity and the generators of the fundamental group to be $l_0,l_t,l_1 $, where each $l_i$,  $i=0,t,1$, encircles only  the pole $i$ once and  $l_0,l_t,l_1 $ are oriented in such a way that
\be\label{eq:ord-mon}
M_0 M_t M_1 M_\infty=\ID,
\ee
where $M_\infty=\exp(2\pi i A_\infty)$. 

\subsection{Riemann-Hilbert correspondence and monodromy manifold}
\label{sec:rh}

Let us denote by  ${\mathcal F}(\theta_0,\theta_t,\theta_1,\theta_\infty)$ the moduli space of rank $2$ meromorphic connection over $\mathbb P^1$ with four simple poles $0,1,t,\infty$ of the form (\ref{eq:fuchs}). Let ${\mathcal M}(a_0,a_t,a_1,a_\infty)$  denote the moduli of monodromy representations $\rho$ up to Jordan equivalence, with the local monodromy data of $a_i$'s
prescribed by (\ref{eq:ai}).
Then the Riemann-Hilbert correspondence 
$$
{\mathcal F}(\theta_1,\theta_2,\theta_3,\theta_\infty)\backslash{\mathcal G}\to {\mathcal M}(\theta_1,\theta_2,\theta_3,\theta_\infty)\backslash GL_2({\mathbb C}),
$$
where $\mathcal G$ is the gauge group \cite{bol}, is defined by associating to each Fuchsian system its monodromy representation class. The representation space 
${\mathcal M}(a_0,a_t,a_1,a_\infty)$ is realised as an affine cubic surface (see \cite{jimbo,iwa}). Let us briefly recall this construction.

With $a_{0t},a_{01},a_{t1}$ defined as in (\ref{eq:aij}),
Jimbo observed that the relation (\ref{eq:ord-mon}) gives rise to the following relation:
$$
a_{0t}^2+a_{01}^2+a_{t1}^2+a_{0t}a_{01}a_{t1}
-\omega_{0t}a_{0t}-\omega_{01}a_{01}-\omega_{t1}a_{t1}+\omega_\infty=0,
$$
with the $\omega$-parameters defined as in (\ref{eq:oij}).
In \cite{iwa}, Iwasaki proved that the tuple 
$(a_0,a_t,a_1,a_{0t},a_{01},a_{t1})$ satisfying the cubic relation (\ref{eq:friecke}) provides a set of coordinates on a large open subset $\mathcal S\subset{\mathcal M}$. In this paper, we restrict to such open set.

\def\thetheorem{B.\arabic{theorem}}
\def\theprop{B.\arabic{prop}}
\def\thelemma{B.\arabic{lm}}
\def\thecor{B.\arabic{cor}}
\def\theexam{B.\arabic{exam}}
\def\theremark{B.\arabic{remark}}
\def\theequation{B.\arabic{equation}}

\appendix{Hamiltonian structure and Okamoto birational transformations}\label{se:bir}

The sixth Painlev\'e equation admits Hamiltonian formulation  \cite{OK1}, i.e. it is equivalent to the following system of first order differential equations:
\be\label{eq:ham}
\dot q = \frac{\partial H}{\partial p},\qquad
\dot p= -\frac{\partial H}{\partial q}
\ee
where $q(t)$ is the solution of the PVI and the Hamiltonian $H(p,q,t)$ is given by the following: 
$$
\begin{array}{ll}
H =& {1\over t(t-1)}\big[q(q-1)(q-t)p^2-\{\theta_0(q-1)(q-t)+\\
&+\theta_1 q(q-t)+(\theta_t-1)q(q-1)\}p+\kappa (q-t)\big],\\
\end{array}
$$
with
$$
\kappa=\frac{(\theta_0+\theta_t+\theta_1-1)^2-(\theta_\infty-1)^2}{4}.
$$
Okamoto studied the group $G_{VI}$ of birational canonical 
transformations of the Hamiltonian system $(p,q,t,H)$, involving different parameters. In 
\cite{OK6} he 
showed that $G_{VI}$ is isomorphic to the extended affine Weyl group of type $F_4$.  Following \cite{NY}, we list Okamoto birational transformations in  the Table here below.

 \begin{center}\begin{tabular}{c||c|c|c|c||c|c|c||}
 & $\theta_0$ &   $\theta_t$  & $\theta_1$  & $\theta_{\infty}$  & $p$ & $q$ & $t$ \\
   \hline\hline
  $s_0$ & $-\theta_0$ &  $\theta_t$  & $\theta_1$  & $\theta_{\infty}$   & $p-\frac{\theta_0}{q}$ &$q$ & $t$ \\
  \hline
  $s_t$ & $\theta_0$ &  $-\theta_t$  & $\theta_1$  & $\theta_{\infty}$  & $p-\frac{\theta_t}{q-t}$  & $q$ & $t$ \\
  \hline
  $s_1$ & $\theta_0$ &  $\theta_t$  & $-\theta_1$  & $\theta_{\infty}$  & $p-\frac{\theta_1}{q-1}$  & $q$ & $t$   \\
  \hline
  $s_{\infty}$ & $\theta_0$ &  $\theta_t$  & $\theta_1$  & $2-\theta_{\infty}$  & $p$ & $q$ & $t$   \\
   \hline
  $s_{\rho}$ & $\theta_0+\rho$ &  $\theta_t+\rho$  & $\theta_1+\rho$  & $\theta_{\infty}+\rho$ & 
    $p$& $q+\frac{\rho}{p}$ & $t$   \\
  \hline\hline
  $r_0$ & $\theta_{\infty}-1$ & $\theta_1$ & $\theta_t$ & $\theta_{0}+1$  &$ \frac{-q(p q +\rho)}{t} $&$t/q$ & $t$  \\
  \hline
  $r_1$ & $\theta_t$  & $\theta_0$ & $\theta_{\infty}-1$& $\theta_1+1$ & $\frac{(q-1)((q-1)p+\rho}{t-1}$ & $\frac{q-t}{q-1}$ & $t$  \\
  \hline
  $r_t$ & $\theta_1$ &  $\theta_{\infty}-1$ &$\theta_0$ & $\theta_t+1$ & $\frac{-(q-t)((q-t)p+\rho)}{t(t-1)} $ & $\frac{t(q-1)}{q-t}$ & $t$   \\
   \hline\hline
  $\sigma_{01}$ & $\theta_1$ & $\theta_t$ & $\theta_0$ & $\theta_{\infty}$  & $-p$ & $1-q$ & $1-t$   \\
  \hline
  $\sigma_{0\infty}$ & $\theta_\infty-1$ & $\theta_t$ & $\theta_1$ & $\theta_0+1$  & $-q(1+\rho+q p)$ 
  & $1/q$ & $1/t$   \\
    \hline
    $\sigma_{0t}$ & $\theta_t$ & $\theta_0$ & $\theta_1$ & $\theta_\infty$  & $-(t-1) p$ 
  & $\frac{t-q}{t-1}$ & $\frac{t}{t-1}$   \\
\hline\end{tabular}\vspace{0.2cm}\\
  \nopagebreak[2] Table: Bi-rational transformations for Painlev\'e VI, 
  $\rho=\frac{2-\theta_0-\theta_t-\theta_1-\theta_\infty}{2}$.
  \vspace{0.2cm}\end{center}

The first five transformations $s_0,s_t,s_1,s_\infty,s_\rho$ generate a group isomorphic to the affine Weyl group $W(D_4^{(1)})$ of type $D_4$. Inaba, Iwasaki and Saito \cite{IIS} proved that the action of this subgroup
is the identity on the monodromy manifold. 
 
The extended affine group of type $D_4$ is obtained by adding the generators $r_0,r_t$ (we listed also $r_1$ even though it is not needed as a generator). These act as permutations on the monodromy matrices and on the monodromy manifold they simply permute and change of two signs as eplained in Subsection \ref{ss:aut} (see for example \cite{LT}).

To obtain the extended affine Weyl group of type $F_4$ we need to add the symmetries $\sigma_{01}, \sigma_{0\infty},\sigma_{0t}$ in  which the time variable is changed by fractional linear transformations. The action on the monodromy matrices was obtained in \cite{DM1}).



\begin{thebibliography}{99}

\footnotesize\itemsep=0pt

\bibitem{bol} 
Bolibruch A.A., The 21-st Hilbert problem for linear Fuchsian systems, {\it Develop-
ments in mathematics: the Moscow school}\/ Chapman and Hall, London, (1993).

\bibitem{ChMa}
Chekhov L.O., Mazzocco M., Isomonodromic deformations and twisted Yangians arising in Teichm\"uller theory, {\it Adv. Math.} {\bf 226} (2010), 4731--4775.

\bibitem{CM} 
Chekhov L.O., Mazzocco M.,
Shear coordinates on the versal unfolding of the $D_4$ singularity,
{\it J. Phys. A. Math. Gen.,}\/ {\bf 43}, (2010), 1--13.


\bibitem{DM}
Dubrovin B.A., Mazzocco, M., 
Monodromy of certain Painlev\'e-VI transcendents and ref\/lection group,
{\it Invent. Math.} {\bf 141} (2000), 55--147.


\bibitem{DM1}
Dubrovin B.A., Mazzocco, M., 
Canonical structure and symmetries of the Schlesinger equations.  
{\it Comm. Math. Phys.,}\/ {\bf 271}  (2007),  no. 2, 289--373. 

\bibitem{EG}
Etingof P., Ginzburg V., Noncommutative del Pezzo Surfaces and Calabi-Yau Algebras,
{\it arXiv:0709.3593v3}, (2007).


\bibitem{fuchs}
Fuchs R., Ueber lineare homogene Differentialgleichungen zweiter Ordnung mit
drei im Endlichen gelegenen wesentlich singul\"aren, {\it Stellen. Math. Ann.,}\/  {\bf 63} (1907)
301--321.

\bibitem{Gar1}
Garnier R., Solution du probleme de Riemann pour les systemes di?\'erentielles
lin\'eaires du second ordre, {\it Ann. Sci. Ecole Norm.. Sup., }\/ {\bf 43} (1926) 239--352.

\bibitem{hit}
Hitchin, N.,
Twistor spaces, Einstein metrics and isomonodromic deformations,
{\it J.Differential Geometry}\/ {\bf 42} (1995), 30--112.

\bibitem{IIS}
Inaba M., Iwasaki K. and Saito M., B\"acklund Transformations of the Sixth Painlev\'e Equation in Terms of Riemann--Hilbert Correspondence, {\it IMRN} {\bf 2004} (2004) no1: 1--30.

 \bibitem{iwa}
 Iwasaki K., An Area-Preserving Action of the Modular Group on Cubic Surfaces and the Painlev VI Equation, {\it Comm. Math. Phys.}\/ {\bf 242} (2003) 185--219.

\bibitem{FG}
Filipuk, G., Gromak, V.~I.,
On the Transformations of the Sixth Painlev\'e Equation,
{\it  J. Nonlinear Math.l Phys.,}\/	 {\bf 10},  (2003), no.2:57--68.	

\bibitem{Gal}
Galbraith. S.~G.,
Mathematics of Public Key Cryptography,
{\it Cambridge University Press}\/ (2011).


\bibitem{jimbo}
Jimbo M., Monodromy Problem and the Boundary Condition for Some Painlev\'e Equations,
{\it Publ. RIMS, Kyoto Univ.,}\/ {\bf 18} (1982) 1137--1161.

\bibitem{JMU} 
Jimbo M., Miwa T. and Ueno K.,
Monodromy preserving deformations of linear ordinary differential
equations with rational coefficients \text{I},
{\it Physica 2D}, {\textbf{2}}, (1981), no. 2, 306--352

\bibitem{MJ1} 
Jimbo M. and Miwa T.,
Monodromy preserving deformations of linear ordinary differential
equations with rational coefficients \text{II},
{\it Physica 2D,}
{\textbf{2}} (1981),
no. 3, 407--448.


\bibitem{kitaev} 
Kitaev, A.,
Quadratic transformations for the sixth Painlev\'e equation,
{\it    Letters in Mathematical Physics,}\/{\bf  21}
(1991), 105--111.

\bibitem{LT}
Lisovyy, O.,  Tykhyy Yu.,
Algebraic solutions of the sixth Painlev\'e equation
{\it arXiv:0809.4873} (2008).

\bibitem{man}
Manin, Yu.I.,
Sixth Painlev\'e equation, universal elliptic curve, and mirror of $\mathbb P^2$,
Geometry of differential equations, 
{\it Amer. Math. Soc. Transl. Ser. 2,} {\bf 186}, Amer. Math. Soc., Providence, RI, (1998),
 131--151.

\bibitem{M}
Mazzocco, M.,
Picard and Chazy solutions to the Painlev\'e VI equation,
{\it Math. Ann.}\/ {\bf 321} (2001) 157--195.



\bibitem{NY}
Noumi M., Yamada Y., A new Lax pair for the sixth Painlev\'e equation associated to $\widehat{\mathfrak{so}}(8)$, {\it Microlocal analysis and complex Fourier analysis,}\/ World Sci. Publ., River Edge, NJ, (2002) 238--252. 



\bibitem{OK1}
Okamoto K.,
\newblock Polynomial Hamiltonians associated with Painlev\'e equations. I. 
\newblock{\em Proc. Japan Acad. Ser. A Math. Sci.} 56 (1980), no. 6, 264--268.

\bibitem{OK6}
Okamoto K.,
\newblock Studies on the Painlev equations. I. Sixth Painlev\'e
equation $P\sb {{\rm VI}}$. 
\newblock{\em  Ann. Mat. Pura Appl.} (4) 146 (1987), 337--381.

\bibitem{RGT}
Ramani, A., Grammaticos, B., and Tamizami, T.,
Quadratic relations in continous and discrete Painlev\'e equations, 
{\it J. Phys. A. Math. Gen.}\/ {\bf 33}, (2000) 3033--3044.

\bibitem{Sch}
Schlesinger L.,
Ueber eine Klasse von Differentsial System Beliebliger
\text{Ordnung} mit \text{Festen Kritischer Punkten},
{\it J. fur Math.,}
{\bf 141}, (1912),
96--145.

\bibitem{silverm}
Silverman, J.H.,
{\em The arithmetic of elliptic curves}, 
Springer, 2009.

\bibitem{OST}
T.~Tsuda, K.~Okamoto and H.~Sakai,
Folding transformations of the Painlev\'e equations, 
{\it Math. Ann.}\/ {\bf 331} (2005) 713--738.

\bibitem{uga}
Ugaglia M., On a Poisson structure on the
space of Stokes matrices, {\it Int. Math. Res. Not.} {\bf 1999} (1999),  no.~9, 473--493,
{http://arxiv.org/abs/math.AG/9902045}{math.AG/9902045}.

\bibitem{vidkit} 
Vidunas. R., Kitaev, A.,
Quadratic transformations of the sixth Painlev\'e equation
with application to algebraic solutions,
{\it  Mathematische Nachrichten,}\/ {\bf  280}
(2007),  1834--1855.

\end{thebibliography}
\end{document}